\documentclass[proof]{pasj01}
\Received{2021/12/17}
\Accepted{2022/05/10}
\onecolumn
\usepackage{natbib}
\usepackage{color}
\usepackage{bm}
\usepackage{lineno}

\newcommand{\Pphoto}{P_{\rm photo}}

\newcommand{\Rs}{R_{\star}}
\newcommand{\As}{A_{\rm spot}}
\newcommand{\bs}{b_{\rm spot}}
\newcommand{\rs}{r_{\rm spot}}
\newcommand{\bsi}{b_{\rm spot,i}}
\newcommand{\ells}{\ell_{\rm s}}
\newcommand{\ello}{\ell_{\rm o}}

\newcommand{\varphis}{\varphi_{\rm s}}
\newcommand{\varphisinit}{\varphi_{\rm s0}}
\newcommand{\varphisiniti}{\varphi_{\rm s0,i}}
\newcommand{\varphio}{\varphi_{\rm o}}
\newcommand{\omegas}{\omega_{\rm s}}
\newcommand{\omegasi}{\omega_{{\rm s},i}}
\newcommand{\thetac}{\theta_{\rm c}}
\usepackage{xspace}
\begin{document}
\title{Analytic model for photometric variation due to starspots on a
  differentially rotating star}
\author{
  Yasushi \textsc{Suto} \altaffilmark{1,2}, 
  Shin \textsc{Sasaki}  \altaffilmark{3}, 
Yuta \textsc{Nakagawa} \altaffilmark{1},
  and Othman \textsc{Benomar} \altaffilmark{2,4}}
\email{suto@phys.s.u-tokyo.ac.jp}
\altaffiltext{1}{Department of Physics, The University of Tokyo, Tokyo
113-0033, Japan}
\altaffiltext{2}{Research Center for the Early Universe, School of
Science, The University of Tokyo, Tokyo 113-0033, Japan}
\altaffiltext{3}{Department of Physics, Tokyo Metropolitan University,
Hachioji, Tokyo 192-0397, Japan}
\altaffiltext{4}{National Astronomical Observatory of Japan,
  Mitaka, Tokyo 181-0015, Japan}

\maketitle

\begin{abstract}
  We present an analytic model of the lightcurve variation for
    stars with non-evolving starspots on a differentially rotating
    surface. The Fourier coefficients of the harmonics of the rotation
    period are expressed in terms of the latitude of the spot,
    $\ells$, and the observer's line-of-sight direction, $\ello$,
    including the limb darkening effect.  We generate different
    realizations of multi-spots according to the model, and perform
    mock observations of the resulting lightcurve modulations. We
    discuss to what extent one can recover the properties of the spots
    and the parameters for the differential rotation law from the
    periodogram analysis.  Although our analytical model neglects the
    evolution of spots on the stellar surface (dynamical motion,
    creation and annihilation), it provides a basic framework to
    interpret the photometric variation of stars, in particular from
    the existing Kepler data and the future space-born mission.  It is
    also applicable to photometric modulations induced by rotation of
    various astronomical objects.
\end{abstract}


\section{Introduction}

The last two decades have seen the birth and growth of the space borne
photometry due to missions like MOST \citep{Walker2003}, CoRoT
\citep{Baglin2006a}, Kepler \citep{Borucki2010} and TESS
\citep{Ricker2014TESS}.  These instruments provided for the first time
long, continuous high-quality photometric data, enabling the detection
of thousands of transiting planets, but also opened a new window on
the dynamic and evolution of stars. For example, one of the most
remarkable achievements of the Kepler space instrument is the
discovery that older low-mass stars rotate too fast compared to
theoretical expectations \citep{Saders2016}. This could only be
established by the combined analysis of the stellar photometric
variability due to spots and to the stellar pulsations \citep[see
  e.g.][]{Kjeldsen1995,JCD1996Science,2010ApJ...713L.169C}. In
general, the study of the rotation-age relation
\citep{Skumanich1972ApJ,Kawaler1988ApJ,MacGregor1991ApJ} is an
important tool to evaluate the age of stars.  The rotation also plays
an important role on the solar and stellar dynamo
\citep{Ossendrijver2003,Varela2016}, itself believed to be important
for sustaining a latitudinal differential rotation.

Observationally, the stellar rotation period can be estimated from a
few independent methods. First, one can combine the equatorial
rotational velocity from Doppler broadening and the stellar
radius. The spectroscopically derived rotation period, however,
depends on the assumed model for the turbulence, and also requires the
values of the stellar radius and inclination that are not
well-determined in general\citep{Kamiaka2018}.  Second, the
asteroseismic analysis of the stellar pulsation can estimate the
rotation period and the stellar inclination simultaneously. The
asteroseismology, however, also required various model
  assumptions in the analysis, and is applicable only to a relatively
small fraction of stars that exhibit measurable oscillations
\citep[e.g.,][]{Appourchaux2008,Huber2013b,Benomar2014b,Lund2017,
  Kamiaka2018,Kamiaka2019}.  Finally, the photometric variation of
lightcurves is by far the most widely used method to estimate the
stellar rotation period, and has been intensively applied for the
Kepler data \citep[e.g.,][]{McQuillan2014,Mazeh2015b,Angus2018}.

The photometric variation is induced by star-spots corotating with the
star.  If those spots do not dynamically evolve on the stellar
surface, it is relatively easy to estimate the stellar rotation
period. In reality, however, the spots have individual lifetimes and
even move on the stellar surface, and stars are not necessarily rigid
rotators \citep[e.g.,][]
{Donati1997,Barnes2005,Donati2010,Roettenbacher2013,Walkowicz2013,Brun2017,
  Benomar2018,Basri2020}.  The formation and dissipation of spots on
the differentially rotating stars, therefore, complicate the
interpretation of the photometrically estimated rotation period
$\Pphoto$. Furthermore, we cannot exclude a possibility that
  that starspots and stellar pulsations may have similar time scales
  in some stellar types, even if not so likely.

Properties of spots have been extensively studied in the past
literature for the Sun
\citep[e.g.,][]{Maunder1904,Zharkov2005,Mandal2021}, and also for
other stars \citep{Morris2020}.  \citet{Roettenbacher2013}, for
instance, achieved a wonderful lightcurve inversion to predict the
starspot evolution on Kepler target KIC 5110407.  Nevertheless, it is
not easy to accurately predict the nature of spots in general.  On the
other hand, the photometric rotation periods combined with the
spectroscopic Doppler broadening have been extensively used to infer
the inclination angle of stars hosting planets
\citep{Sanchis-Ojeda2011a,Sanchis-Ojeda2011b,Hirano2012,
  Louden2021,Albrecht2021}, which have profound implications for the
spin-orbit architecture of exoplanetary
systems\citep{Queloz2000,Ohta2005,SS2021}.  Therefore, it is still
useful to have parameterized templates for the photometric variation
of stellar lightcurves due to non-evolving starspots. This is the
purpose of this paper. We present an analytic model of the photometric
lightcurves induced by starspots on a differentially rotating stellar
surface {\it assuming that they do not evolve during the finite
  observing duration}. We compute mock lightcurves based on the
multi-spot model, and address how to interpret the measured
distribution of peaks in the Lomb-Scargle periodogram in terms of the
stellar differential rotation law.

The rest of the paper is organized as follows. We derive the
photometric variation pattern due to a single infinitesimal spot in
section 2.  The resulting lightcurve modulation including the limb
darkening effect is expressed in the Fourier series expansion.  In
section 3, we apply the analytic model for multispots on a
differentially rotating star.  Then we generate simulated lightcurves
adopting the statistical distribution model of the Sun spots,
perform the Lomb-Scargle analysis, and examine the information content
of the resulting power spectra. Final section is devoted to summary
and conclusion of the paper.  The Fourier expansion coefficients in
our analytic model are given in Appendix.

\section{Photometric variation due to a single starspot}

\begin{figure}
 \begin{center}
 \includegraphics[width=10cm, bb = 0 0 723 451]{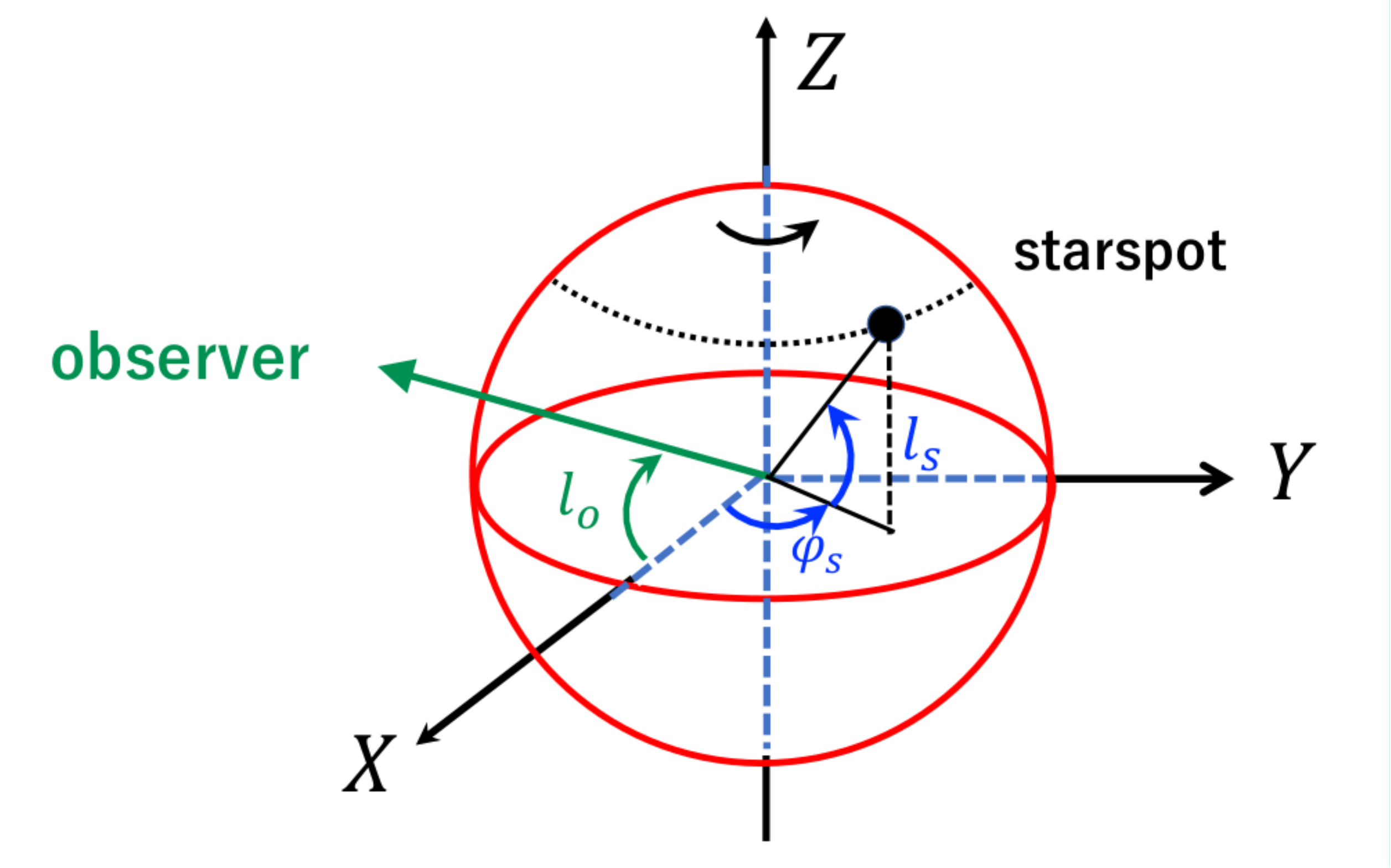} 
 \end{center}
 \caption{Schematic illustration of the observer and a starspot. The
   stellar rotation axis is chosen to be the $Z$-axis. The
     location of the spot is specified by the latitude $\ells$ and the
     azimuth angle $\varphis$, and the direction of the observer's
     line-of-sight is defined by the latitude $\ello$ on the $X-Z$
     plane ($\varphi=0$).}
 \label{fig:obs-spot}
\end{figure}

As illustrated in Figure \ref{fig:obs-spot}, we consider a spherical
star with radius $\Rs$, and parameterize a position vector on
the stellar surface in terms of its latitude $\ell$ and
longitude $\varphi$:
\begin{eqnarray}
  \label{eq:rspot}
  \bm{r}_{\star} =
  \left( \begin{array}{c} x_{\star} \\ y_{\star} \\ z_{\star}
  \end{array} \right)
  = \Rs
\left( \begin{array}{c}
 \cos \ell \cos\varphi \\
 \cos \ell \sin\varphi\\
 \sin \ell \end{array} \right) ,
\end{eqnarray}
where the $z$-axis is chosen to be the direction of the stellar
rotation.  In what follows, we assume that the surface angular
  velocity $\omega(\ell)$ at $\ell$ is given
  by the following parameterized model for the
  latitudinal differential rotation:
\begin{equation}
\label{eq:diff-law}
  \omega (\ell) = \omega_0 (1 - \alpha_2 \sin^2\ell -
\alpha_4 \sin^4\ell).
\end{equation}
For the Sun, $\omega_{0\odot}\approx 2.972\times
10^{-6}\mathrm{~rad~s^{-1}}$, $\alpha_{2\odot} \approx 0.163$, and
$\alpha_{4\odot} \approx 0.121$ \citep{1990ApJ...351..309S}.
Thus, the angular velocity at $\ell=30^\circ$ is about 5 percent
  smaller than its equatorial value.

Without loss of generality, we consider a distant observer located at
$\varphio=0$ and $\ello$. Thus the unit vector toward the observer is
\begin{equation}
  \label{eq:eo}
  {\bm e}_{\rm o}=(\cos\ello, 0, \sin\ello).
\end{equation}
According to equation
(\ref{eq:diff-law}), the longitude of the starspot located at the
latitude $\ells$  at epoch $t$ becomes
\begin{equation}
\label{eq:varphis-t}
\varphis(t) = \varphi_{\rm s0} +\omegas t
\end{equation}
due to the stellar surface rotation, where $\varphi_{\rm s0}$
  is the longitude at which the spot is located on the stellar surface
  initially ($t=0$), and $\omegas \equiv \omega(\ells)$ is
  the angular velocity of the spot at $\ells$ defined as equation
  (\ref{eq:diff-law}).

\subsection{A single infinitesimal starspot without limb darkening
\label{subsec:singlespot-wo-ld}}

A normalized lightcurve of the stellar surface is
\begin{equation}
\label{eq:lightcurve}
L(t) = \frac{\displaystyle
  \int \cos\ell d\ell d\varphi ~ A(\ell, \varphi)
      K(\ell, \varphi; \ello, \varphio)}
{\displaystyle \int \cos\ell d\ell d\varphi
  ~ K(\ell, \varphi; \ello, \varphio)}.
\end{equation}
where the integration is over the stellar surface, $A$ indicate the
surface intensity distribution, and $K$ is the weighting kernel of the
surface visible for the observer
\citep[e.g.,][]{Fujii2010,Fujii2011,Farr2018,Haggard2018,Nakagawa2020}.

In the case of a single infinitesimal starspot at $(\ells,
  \varphis)$ on a homogeneous sphere,  we set
\begin{equation}
\label{eq:intensity-spot}
A(\ell, \varphi) = 1 - \frac{\bs}{\cos\ell}
\delta(\ell-\ells, \varphi-\varphis).
\end{equation}
The dimensionless parameter $\bs$ represents the amplitude of the
photometric modulation \citep{Dorren1987,Haggard2018}.

Sunspots consist of the central darker part ({\it umbra}), and the
surrounding lighter part ({\it penumbra}). In addition, there is a
type of brighter spots ({\it faculae}).  Note that it is not
  necessary to specify separately the temperature and area of spots,
  since the amplitude of the photometric variations of stars depends
  on their flux (i.e., $\bs$) alone. As described in \S
  \ref{subsec:mock-LS}, we consider the spot distribution directly
  derived from the observed properties using the photometric variation
  data of the Sun.

 Our analytic formulation, however, is general and applicable to
 various types of spots including umbra, penumbra and faculae
 ($\bs<0$), if we  employ their distribution function properly.

Without loss of generality, we can define the initial phase of the
single starspot to be $\varphis(t=0)=\varphi_{\rm s0}=0$. For an
isotropically emitting stellar surface, the weighting kernel $K$ is
equivalent to the visibility computed from the direction cosine $\mu
\equiv {\bm e}_{\rm o}\cdot{\bm e}_{\star}$ between the stellar
surface $\bm{r}_\star=\Rs \bm{e}_\star$ and the observer ${\bm e}_{\rm
  o}$. The spot is visible (invisible) to the observer if $\mu>0$
  ($\mu<0$).  Thus the weighting kernel is simply computed from
  equations (\ref{eq:rspot}) and (\ref{eq:eo}) as
\begin{eqnarray}
K(\ell, \varphi; \ello, \varphio)
&=& \max (\mu, 0 ) \cr
&=& \max ( \cos\ello \cos\ell \cos\varphi+\sin\ello\sin\ell, 0 )\cr
\label{eq:kernel}
&=& \cos\ello \cos\ell
\max(\cos\varphi + \tan\ello \tan\ell, 0).
\end{eqnarray}

Substituting equations (\ref{eq:intensity-spot}) and
  (\ref{eq:kernel}) into equation (\ref{eq:lightcurve}), one obtains a
  normalized lightcurve modulation due to a single starspot on an
  otherwise homogeneous spherical surface:
\begin{equation}
\label{eq:Lt-spot}
L_{\rm s}(t) \equiv L(t)-1 = - \frac{\bs \cos\ello \cos\ells}{\pi} 
\max(\tan\ello \tan\ells + \cos\varphis(t), 0).
\end{equation}
For $\ello=\ells=\varphis(t)=0$, equation (\ref{eq:Lt-spot}) reduces
to $L_{\rm s}(t) = - \bs/\pi$, and the denominator $\pi$ indeed
corresponds to the visible projected area of the stellar surface $\pi
\Rs^2$. Thus, we note that $\bs$ represents the {\it effective area}
of the spot $\As$ in units of $\Rs^2$, instead of $\pi \Rs^2$.  Since
$\bs$ in equation (\ref{eq:intensity-spot}) is defined with respect to
the flux, $\As$ is equivalent to the geometric area of the spot $S$
only when it is completely black.  In general, $\As$ should be
interpreted to represent an flux-weighted area of the spot.

In what follows, we adopt a parameterized model of the
  distribution of $\As$ that is directly estimated from the observed
  photometric variations of the Sun (see \S
  \ref{subsec:mock-LS}). Then we will compute the dimensionless parameter
  $\bs \equiv \As/\Rs^2$.  If the black-body approximation for the
  stellar surface and the spot is valid, the effective and geometric
  areas of the spot are related as $\As \approx (1-T_{\rm
    spot}^4/T_\star^4)S$ with $T_\star$ and $T_{\rm spot}$ being the
  temperatures of the star and the spot.

Equation (\ref{eq:Lt-spot}) indicates that the starspot is visible at
$t$ if
\begin{eqnarray}
\label{eq:visible-criterion}
     \cos\varphis(t) + \tan\ello \tan\ells  > 0 .
\end{eqnarray}
For convenience, let us introduce a parameter
\begin{eqnarray}
\label{eq:def-gamma}
  \Gamma \equiv \tan\ello \tan\ells.
\end{eqnarray}
A starspot with $\Gamma>1$ is always visible
  to the observer, and equation (\ref{eq:Lt-spot}) reduces to
\begin{equation}
\label{eq:Lt-spot-visible}
L_{\rm s}(t) = - \frac{\bs \cos\ello \cos\ells}{\pi} 
\left[\cos\varphis(t)+\Gamma\right].
\end{equation}
If $\Gamma < -1$, on the other hand, the starspot is totally invisible
and $L_{\rm s}(t)=0$.

A starspot with $|\Gamma| \leq 1$ becomes visible periodically as
the stellar surface rotation. In this case, equation
(\ref{eq:Lt-spot}) is expanded analytically in the Fourier series.
The result is
\begin{eqnarray}
\label{eq:Lt-spot-Fourier}
L_{\rm s}(t) = - \frac{\bs \cos\ello \cos\ells}{\pi^2}
&&\Big[
  \left(\sin\thetac-\thetac\cos\thetac\right)
  +\left(\thetac -\sin\thetac\cos\thetac\right)\cos\omegas t\cr
&&  + \sum_{n=2}^\infty
\left(\frac{\sin (n-1)\thetac}{n(n-1)}
-\frac{\sin (n+1)\thetac}{n(n+1)}\right)\cos n\omegas t
\Big],
\end{eqnarray}
where the parameter $\thetac$ is defined through $\Gamma =
  \tan\ello \tan\ells \equiv -\cos\thetac (0<\thetac<\pi)$; see
  Appendix \ref{sec:FT-ld} for the derivation of equation
  (\ref{eq:Lt-spot-Fourier}).

\begin{figure}
 \begin{center}
 \includegraphics[width=8cm, bb = 0 0 576 576]{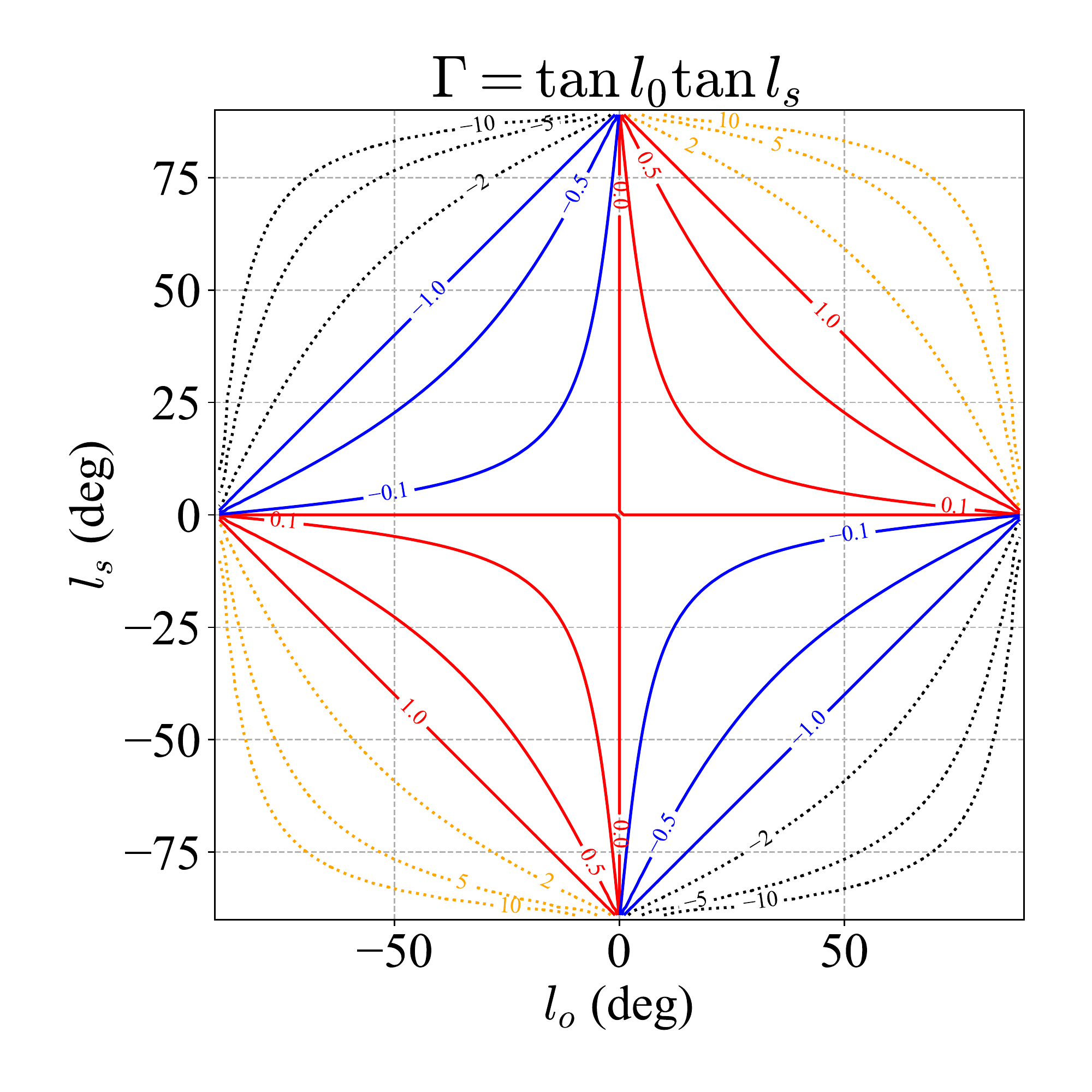} 
 \includegraphics[width=8cm, bb = 0 0 576 576]{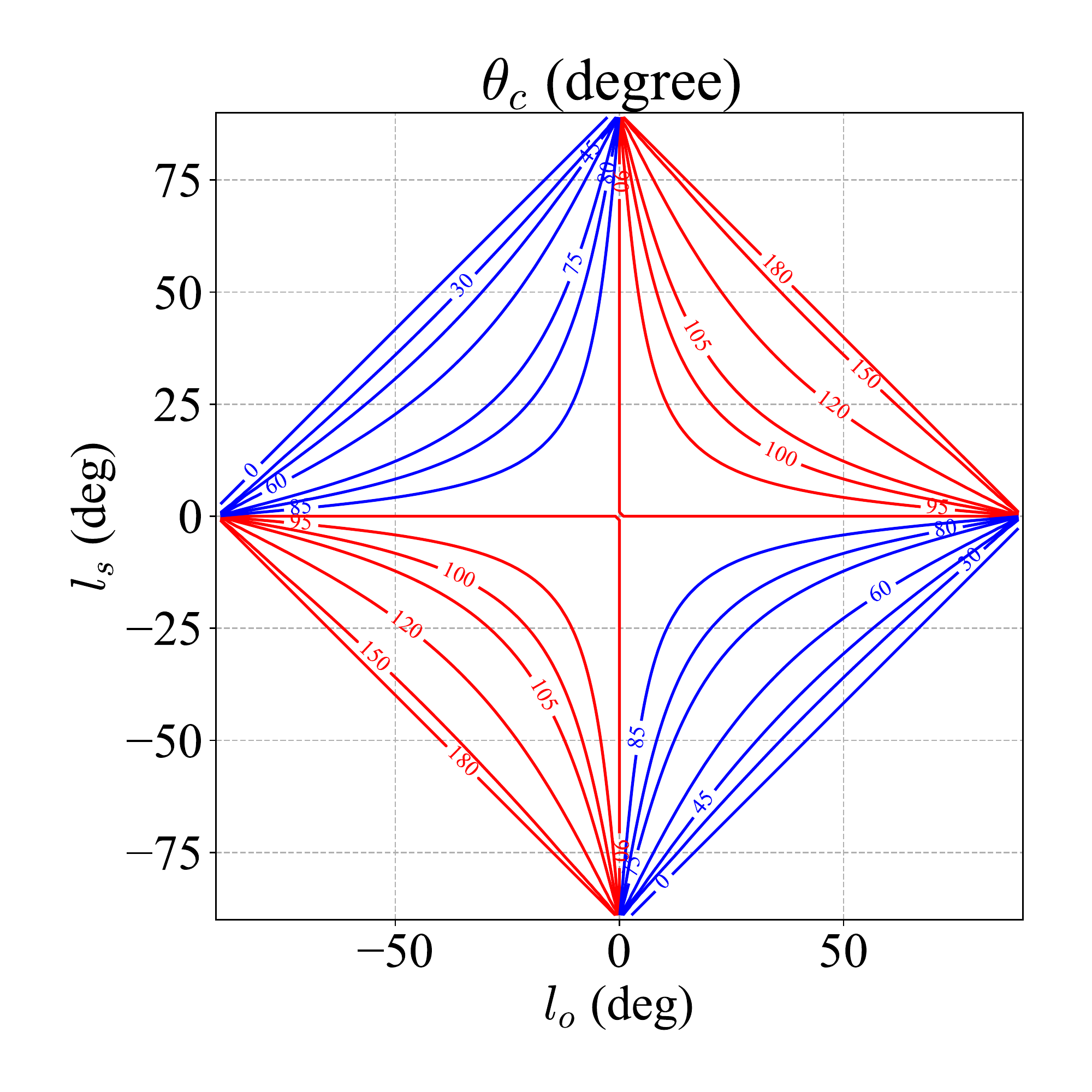} 
 \end{center}
 \caption{Contours of $\Gamma (\equiv \tan\ello \tan\ells)$ and
   $\thetac (\equiv - \cos^{-1} \Gamma$ for $|\Gamma|<1)$
   on $\ello$ -- $\ells$ plane. Red and blue solid lines in both
   panels indicate contours for $0 \leq \Gamma \leq 1$ ($90^\circ \leq
   \thetac \leq 180^\circ$) and $-1 \leq\Gamma <0$ ($0^\circ \leq
   \thetac < 90^\circ$), for which the corresponding starspot
     becomes visible periodically to the observer as the star
     rotates. The orange and black lines correspond to those spots
     that are always visible and invisible to the observer,
     respectively.} 
 \label{fig:gamma-thetac}
\end{figure}

The visibility of a single starspot is determined by the
  parameter $\Gamma$ or equivalently $\thetac$. We plot the contours
  of $\Gamma$ and $\thetac$ on $\ello$ -- $\ells$ plane in the left
  and right panels of Figure \ref{fig:gamma-thetac}, respectively.
  For a roughly edge-on view observer ($|\ello| \ll 1$), spots located
  near the equatorial plane ($|\ells| \ll 1$) correspond to $|\Gamma|
  \approx |\ello\ells| \ll 1$, and $\thetac \approx \pi/2+\ello\ells$.

\subsection{A single infinitesimal starspot with limb darkening}

The stellar limb darkening produces an additional modulation to the
photometric variation due to the starspot.  Adopting the quadratic
limb darkening law, the normalized stellar surface intensity at
$\bm{r}_\star$ is characterized by the two limb darkening parameters
$u_1$ and $u_2$ as
\begin{eqnarray}
I(\mu)=1-u_1(1-\mu)-u_2(1-\mu)^2 = (1-u_1-u_2) +(u_1+2u_2)\mu - u_2\mu^2,
\end{eqnarray}
where $\mu= {\bm e}_{\rm o}\cdot{\bm e}_{\star}$ is the direction
  cosine that we defined before. We adopt the values of $u_1$ and
  $u_2$ from the Sun \citep{Cox2000book}: $u_1 = 0.47$ and $u_2 =
  0.23$ at $550$ nm (they become $0.42$ and $0.23$, respectively, at
  $600$ nm).

Including the limb darkening effect, equation
  (\ref{eq:lightcurve}) is generalized to be
\begin{equation}
\label{eq:lightcurve-ld}
L(t) = \frac{\displaystyle
  \int \cos\ell d\ell d\varphi ~ I(\mu) A(\ell, \varphi)
      K(\ell, \varphi; \ello, \varphio)}
{\displaystyle \int \cos\ell d\ell d\varphi
  ~ I(\mu) K(\ell, \varphi; \ello, \varphio)}.
\end{equation}
Since the denominator of equation (\ref{eq:lightcurve-ld}) is
\begin{eqnarray}
&&\int \cos\ell d\ell d\varphi
~ I(\mu) K(\ell, \varphi; \ello, \varphio)
= \int_0^{2\pi}d\phi\int_0^1 \mu d\mu \left[1 - u_1(1-\mu)
- u_2 (1-\mu)^2\right]\cr
=&& 2\pi\left(\int_0^1\mu d\mu - u_1 \int_0^1\mu(1-\mu) d\mu
- u_2 \int_0^1\mu(1-\mu)^2 d\mu\right)\cr
=&& \pi\left(1 - \frac{u_1}{3} - \frac{u_2}{6}\right),
\end{eqnarray}
equation (\ref{eq:Lt-spot}) is now written as
\begin{equation}
\label{eq:Lt-spot-ld}
L_{\rm s}(t) = - \frac{\bs}{\pi}
\left(1 - \frac{u_1}{3} - \frac{u_2}{6}\right)^{-1}
\max(\mu_{\rm s}, 0) I(\mu_{\rm s}).
\end{equation}
where $\mu_{\rm s}= \cos\ello \cos\ells(\cos\omegas t + \Gamma)$.

Similarly to the previous subsection, equation (\ref{eq:Lt-spot-ld})
for $|\Gamma|\le 1$ is expanded analytically in the Fourier series.
The derivation is explicitly given in Appendix \ref{sec:FT-ld},
  and the normalized lightcurve modulation including the limb
  darkening effect is summarized in the following expression:
\begin{eqnarray}
\label{eq:FTLt-spot-ld}
L_{\rm s}(t) = - \frac{\bs}{\pi}
\left(1 - \frac{u_1}{3} - \frac{u_2}{6}\right)^{-1}
\times \left\{
\frac{A_0}{2} + \sum_{n=1}^{\infty} A_n \cos n \omegas t \right\},
\end{eqnarray}
where
\begin{equation}
\label{eq:A-Gamma<1}
A_n \equiv  (1 - u_1 - u_2) (\cos \ello \cos \ells) a_n + (u_1 + 2 u_2) 
(\cos \ello \cos \ells)^2 b_n - u_2 (\cos \ello \cos \ells)^3 c_n,
\end{equation}
and the coefficients $a_n$, $b_n$, and $c_n$ are explicitly given in
Appendix \ref{sec:FT-ld}.

For spots with $\Gamma>1$, $\mu_{s}$ is always positive, and the
corresponding lightcurve is written in the same form as equation
(\ref{eq:Lt-spot-ld}) by replacing $A_n$ by $\tilde{A}_n$, which are
given in Appendix \ref{sec:FT-ld-2}.

Figures \ref{fig:ello0}, \ref{fig:ello30}, and \ref{fig:ello60} show
the trajectories of a single spot on a rotation stellar surface and
the corresponding normalized lightcurves $L_s(t)/\bs$ against
$t/P_{\rm spin}(\ells)$ for an observer located at $\ello=0^\circ$,
$30^\circ$, and $60^\circ$, respectively.  Black, red, blue and orange
curves indicate the results for the spot at the latitude of
$\ells=0^\circ$, $15^\circ$, $45^\circ$, and $75^\circ$.  Solid and
dashed lines in the right panels indicate the lightcurves with and
without limb-darkening (LD).

The right panel of Figure \ref{fig:ello0} shows that the modulation
amplitude $|L_s(t)/\bs|$ without the limb darkening effect becomes
$1/\pi$ for $\ello=\ells=0^\circ$ (black-dashed curve) at
$\varphis(t)=0$.  Limb darkening decreases the effective visible
  area of the entire surface by a factor of $(1 - u_1/3 -
  u_2/6)^{-1}$, while that of the starspot by a factor of $I(\mu_{\rm
    s})$. Depending on the location of the spot, $\ello$,
  $\ells$, and $\varphis(t)$, the resulting $|L_s(t)/\bs|$ with limb
  darkening becomes either smaller or larger than that without limb
  darkening; see Figures \ref{fig:ello0}, \ref{fig:ello30} and
  \ref{fig:ello60}.

\begin{figure}
 \begin{center}
 \includegraphics[width=6cm, bb = 0 0 432 432]{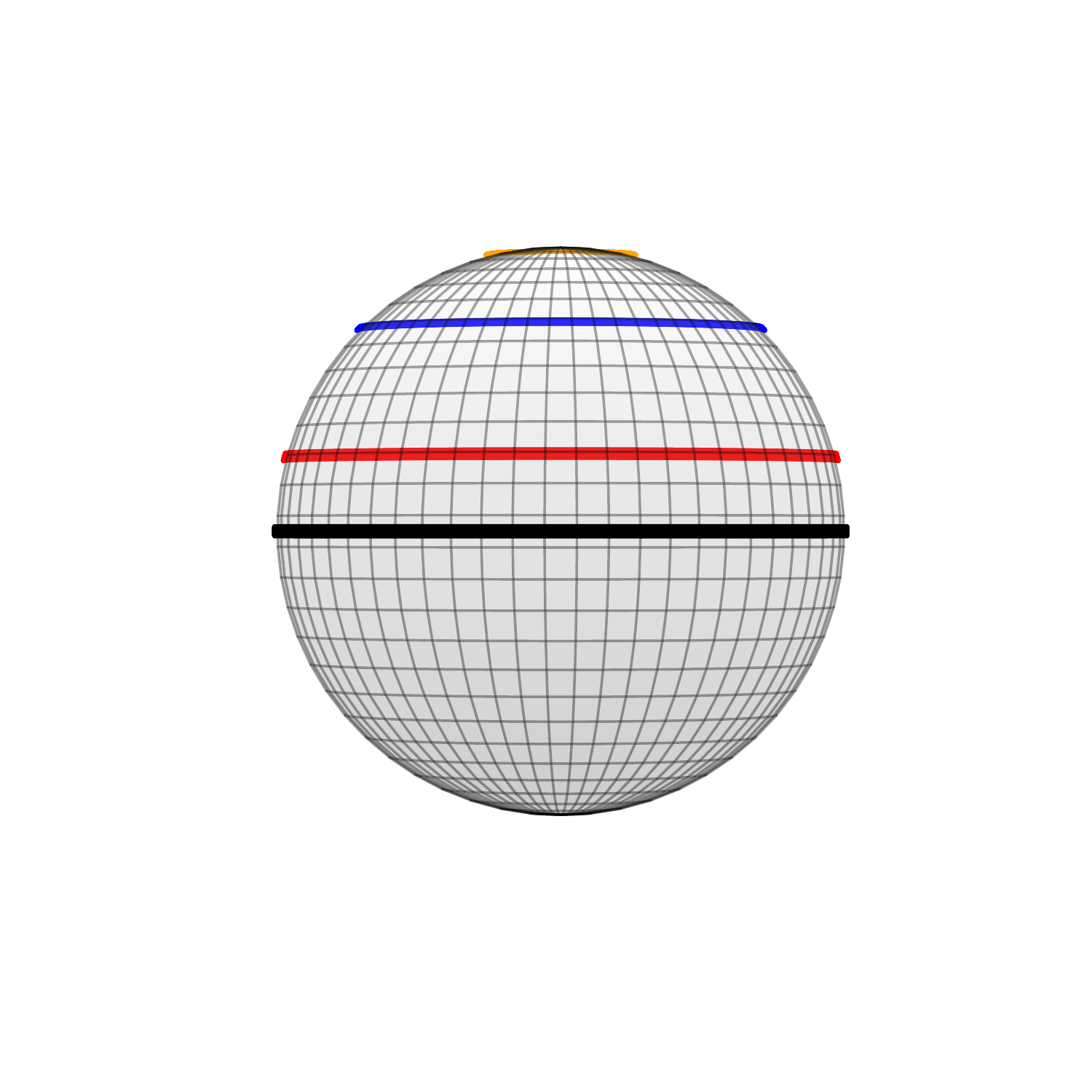} 
 \includegraphics[width=9cm, bb = 0 0 670 531]{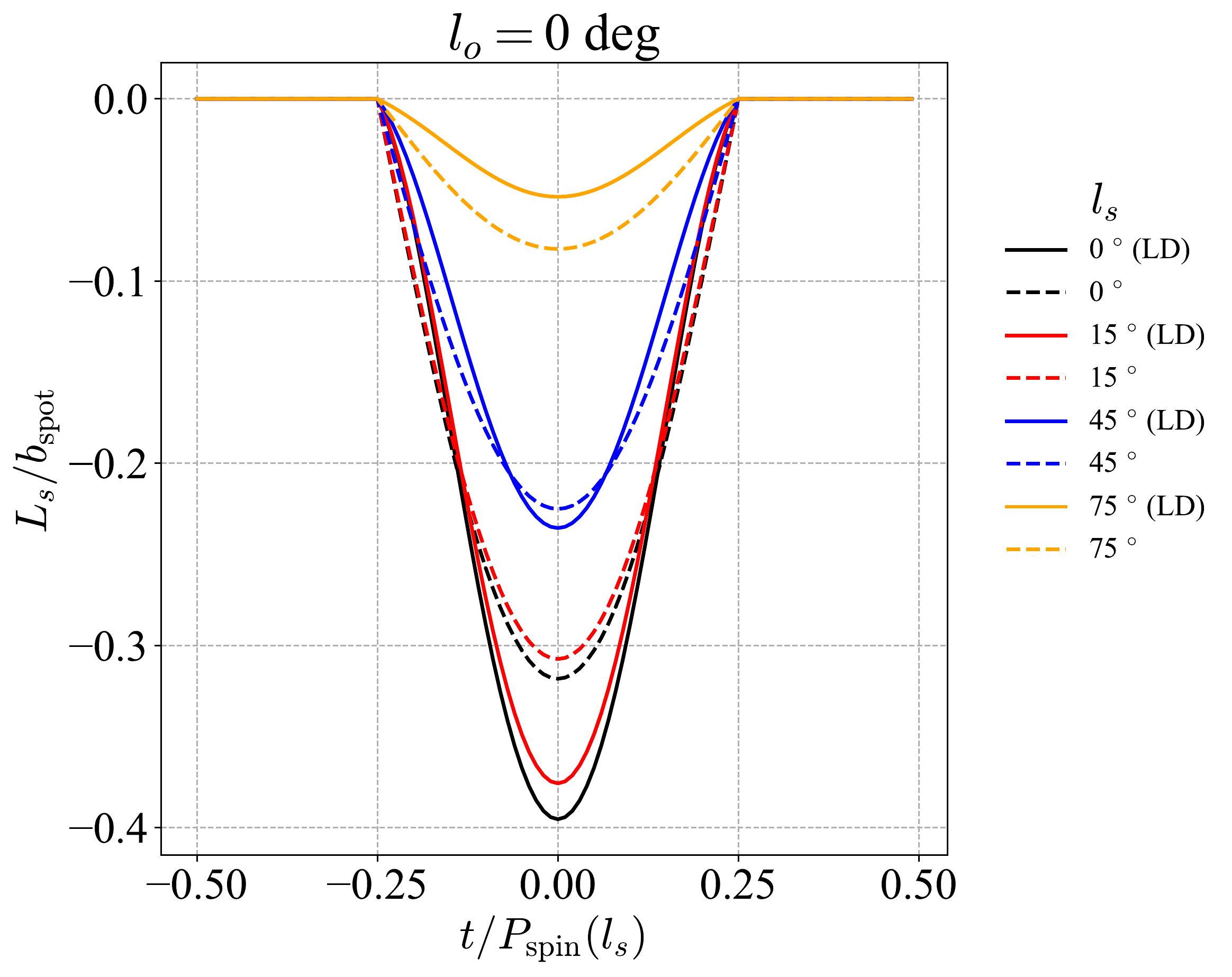} 
 \end{center}
 \caption{Photometric modulation due to a single spot at $\ells$
     viewed from the observer at $\ello=0^\circ$.  {\it left:}
     trajectories of four spots at $\ells=0^\circ$ (black), $15^\circ$
     (red), $45^\circ$ (blue), and $75^\circ$ (orange) on the stellar
     surface. {\it right:} modulation curves $L_s/\bs$ over the one
     rotation period of each spot $P_{\rm spin}(\ells)$ in the left
     panel.  Solid and dashed lines correspond to those with and
     without limb darkening (LD) for the differential rotation
     parameters of $\alpha_2=\alpha_{2\odot}$ and
     $\alpha_4=\alpha_{4\odot}$; see equation (\ref{eq:diff-law}).}
 \label{fig:ello0}
\end{figure}
\begin{figure}
 \begin{center}
 \includegraphics[width=6cm, bb = 0 0 432 432]{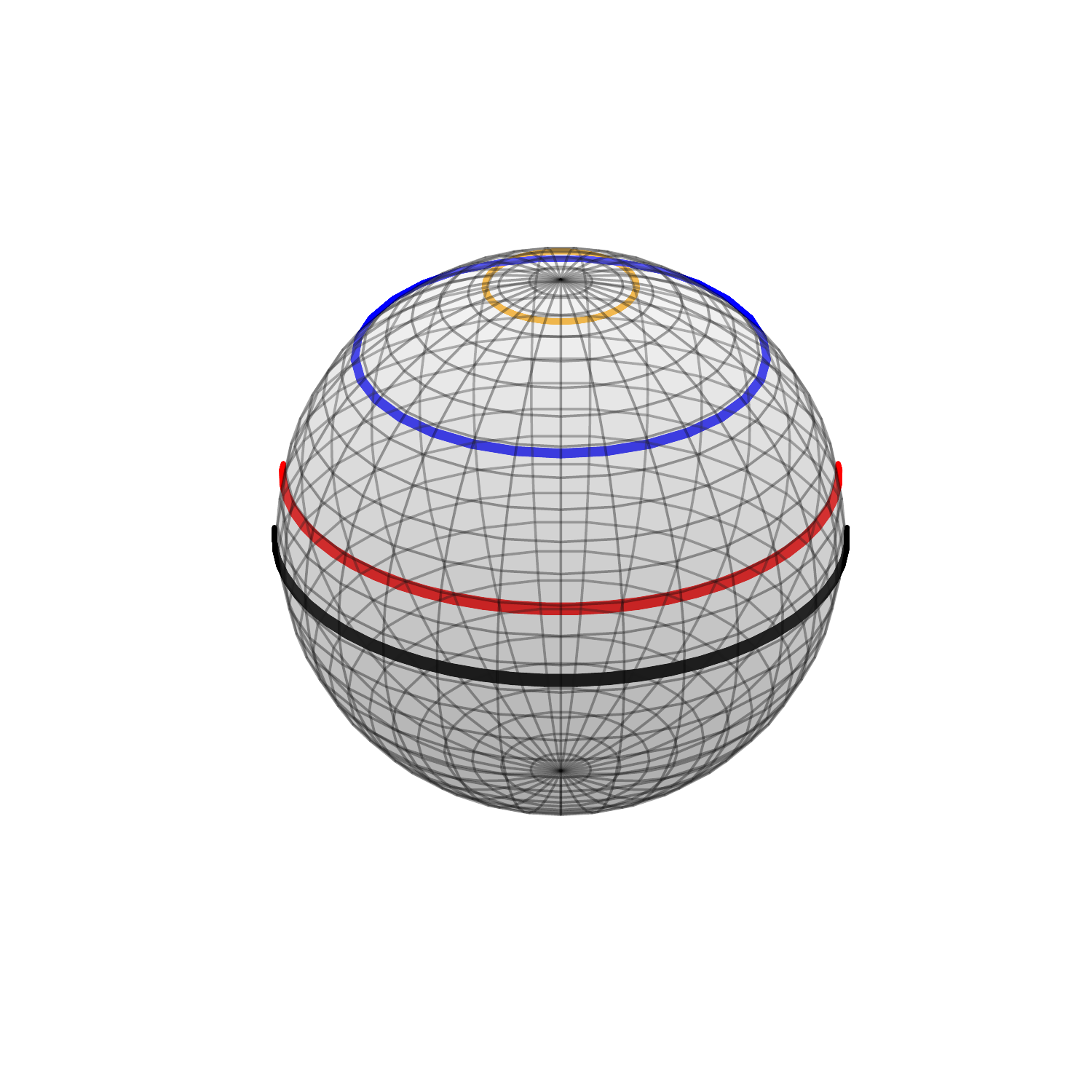} 
 \includegraphics[width=9cm, bb = 0 0 670 531]{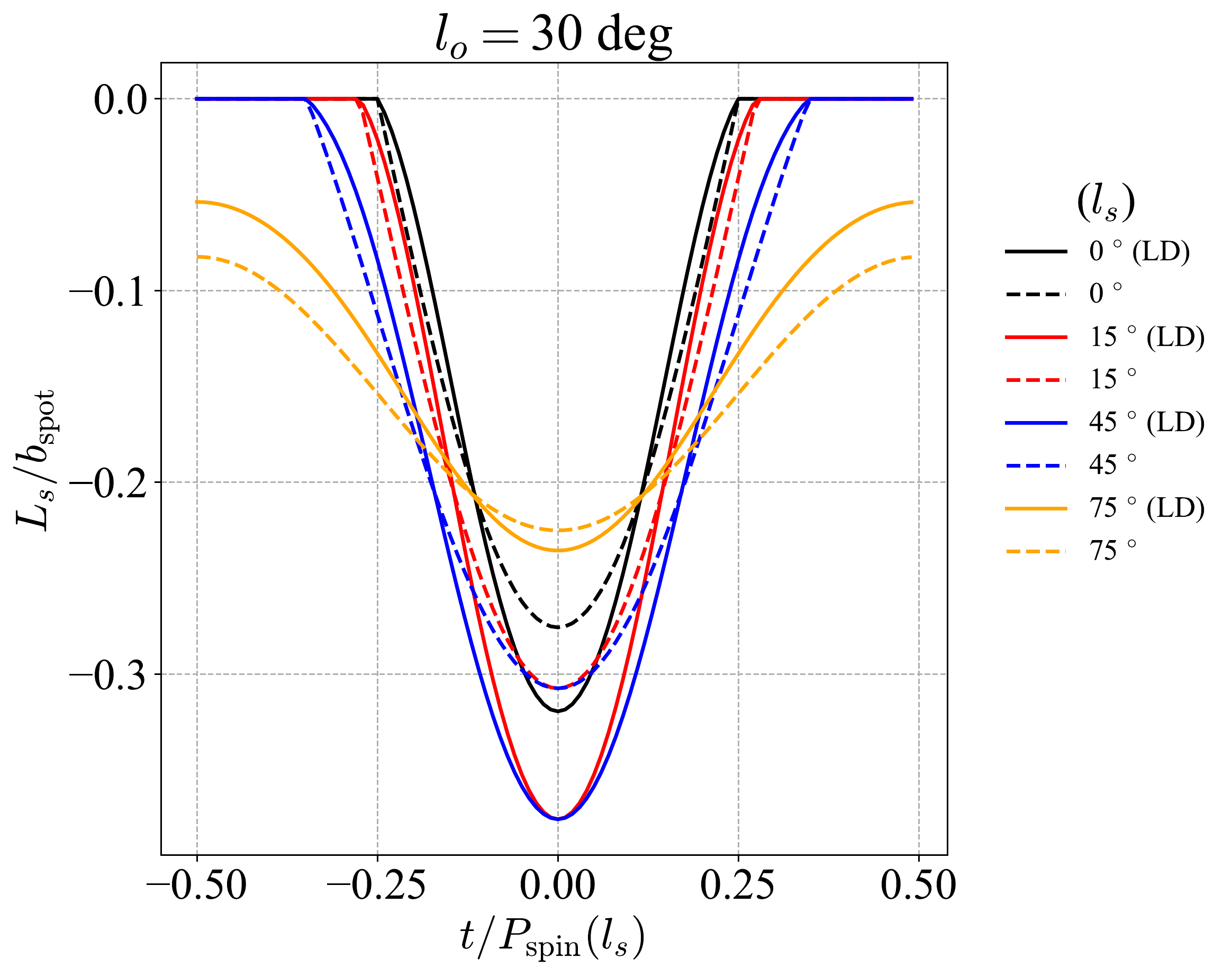} 
 \end{center}
 \caption{Same as Figure \ref{fig:ello0} but viewed from the observer
   at $\ello=30^\circ$.}
  \label{fig:ello30}
\end{figure}
\begin{figure}
 \begin{center}
 \includegraphics[width=6cm, bb = 0 0 432 432]{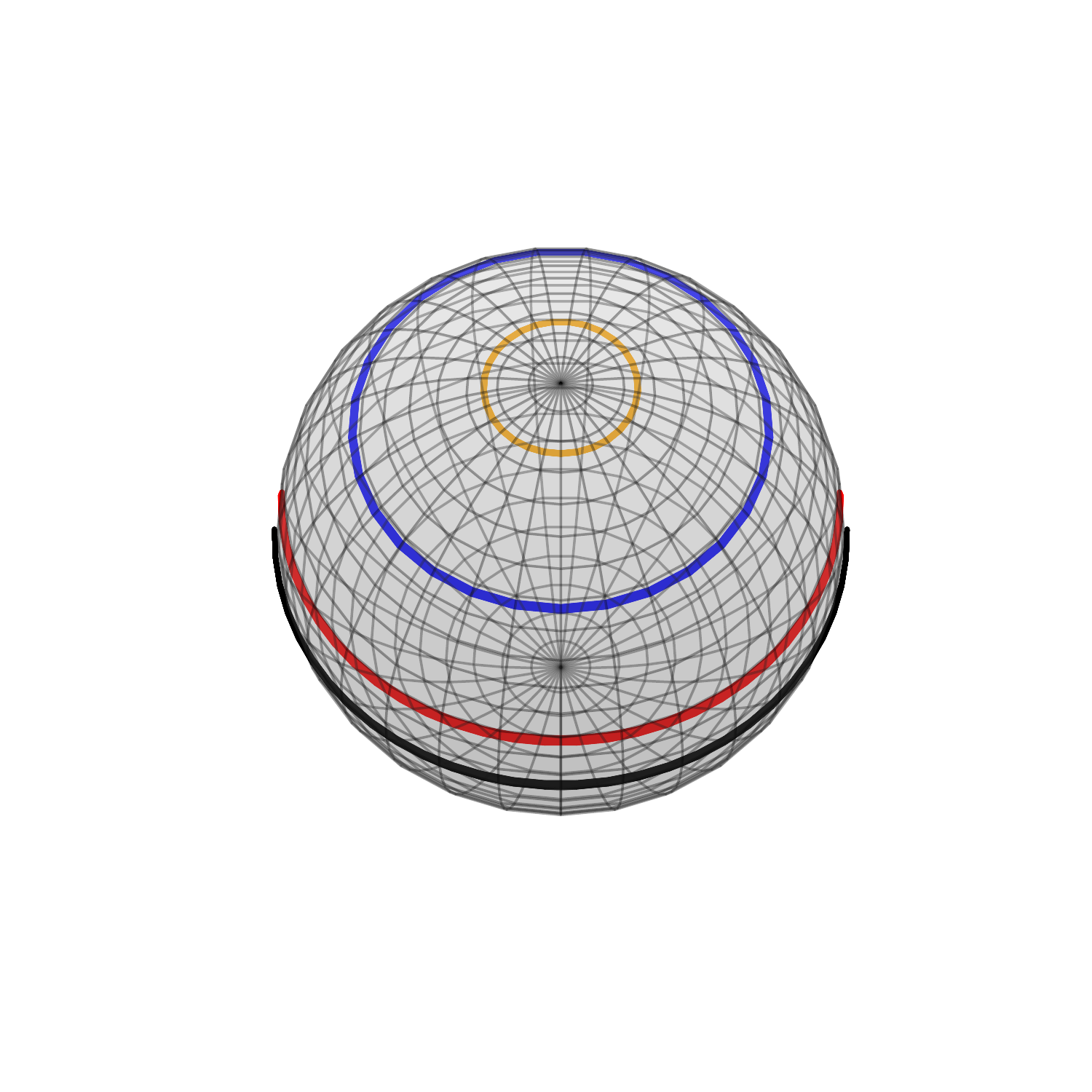} 
 \includegraphics[width=9cm, bb = 0 0 670 531]{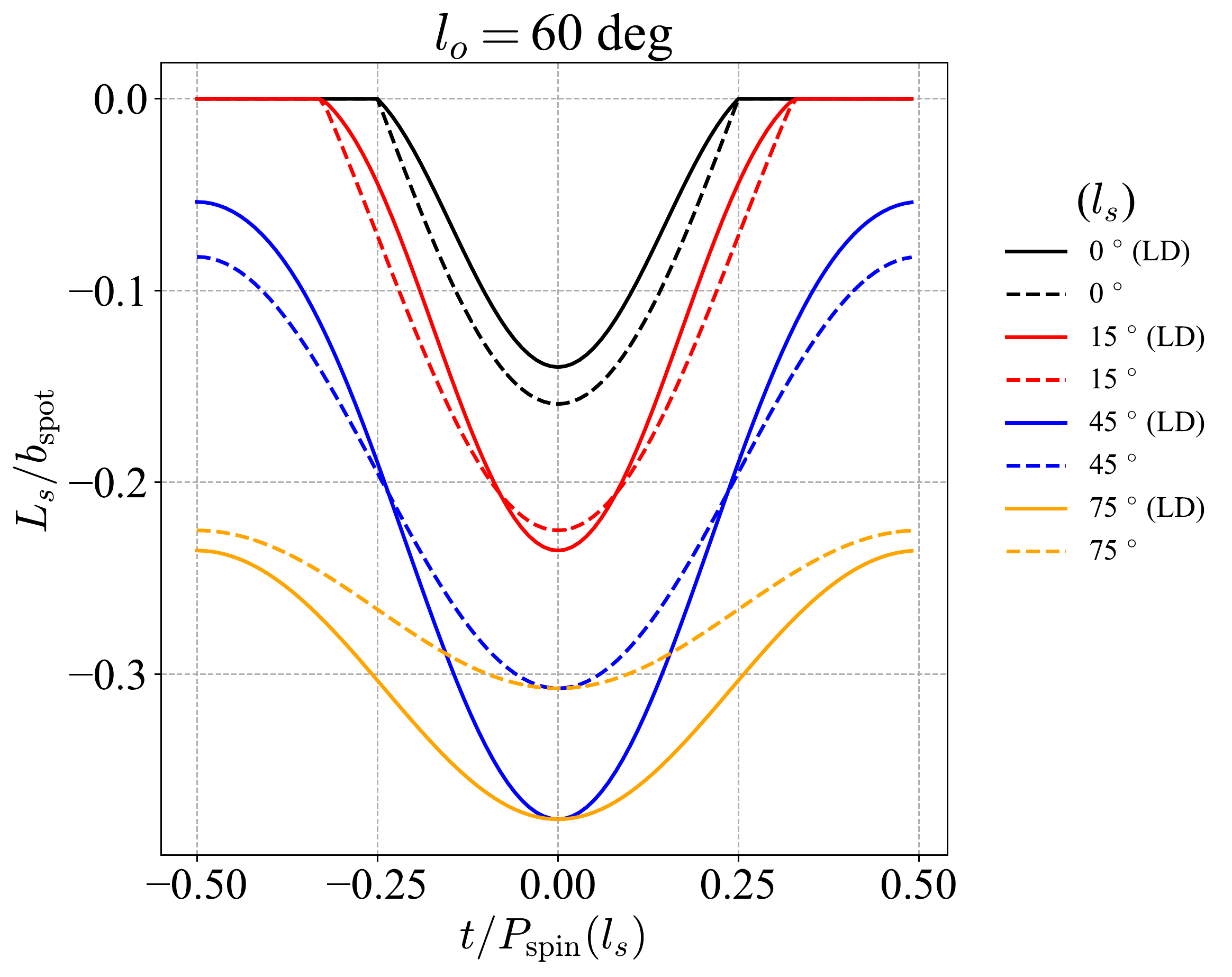} 
 \end{center}
 \caption{Same as Figure \ref{fig:ello0} but viewed from the observer
   at $\ello=60^\circ$.}
  \label{fig:ello60}
\end{figure}

\begin{figure}
 \begin{center}
 \includegraphics[width=8cm, bb = 0 0 576 576]{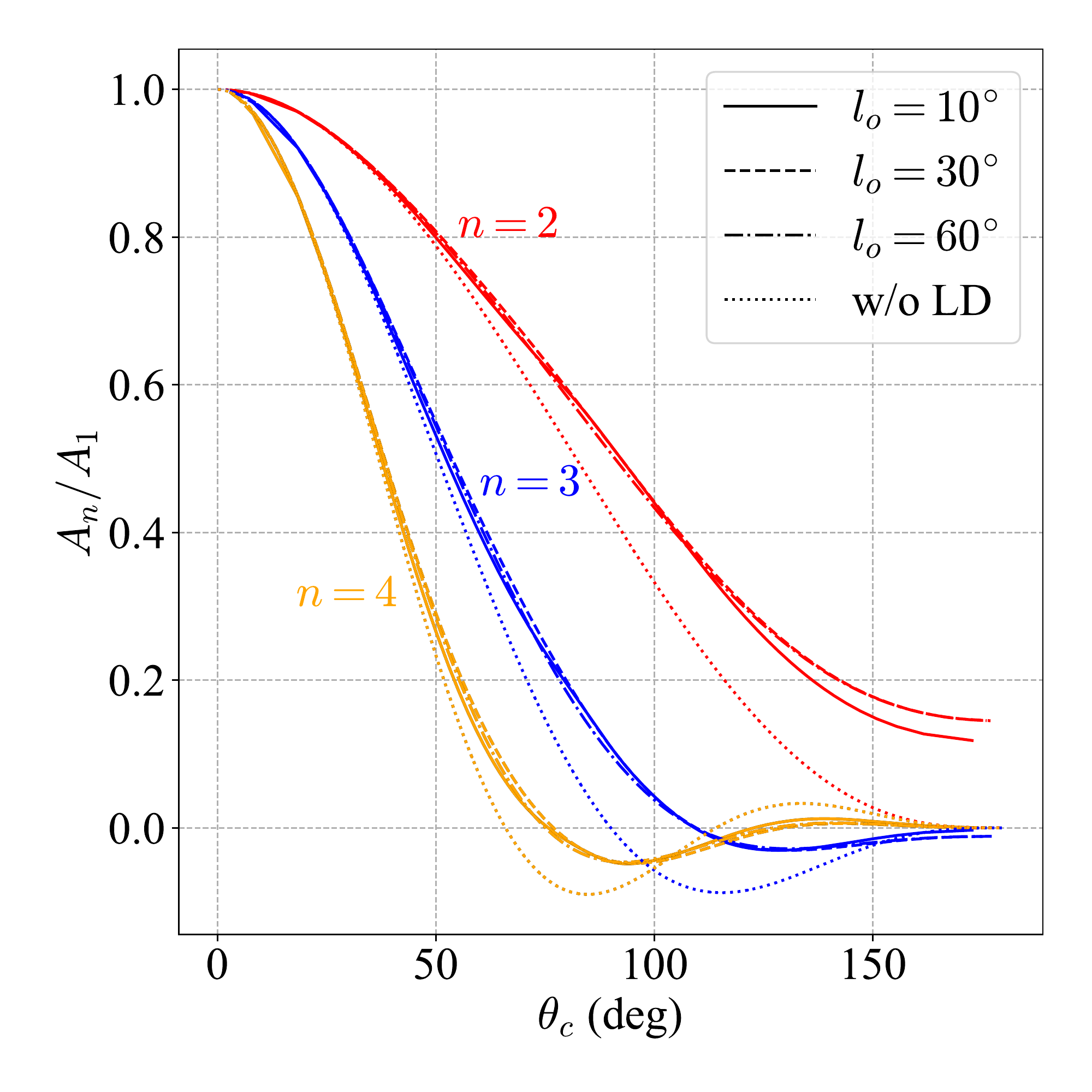} 
 \includegraphics[width=8cm, bb = 0 0 576 576]{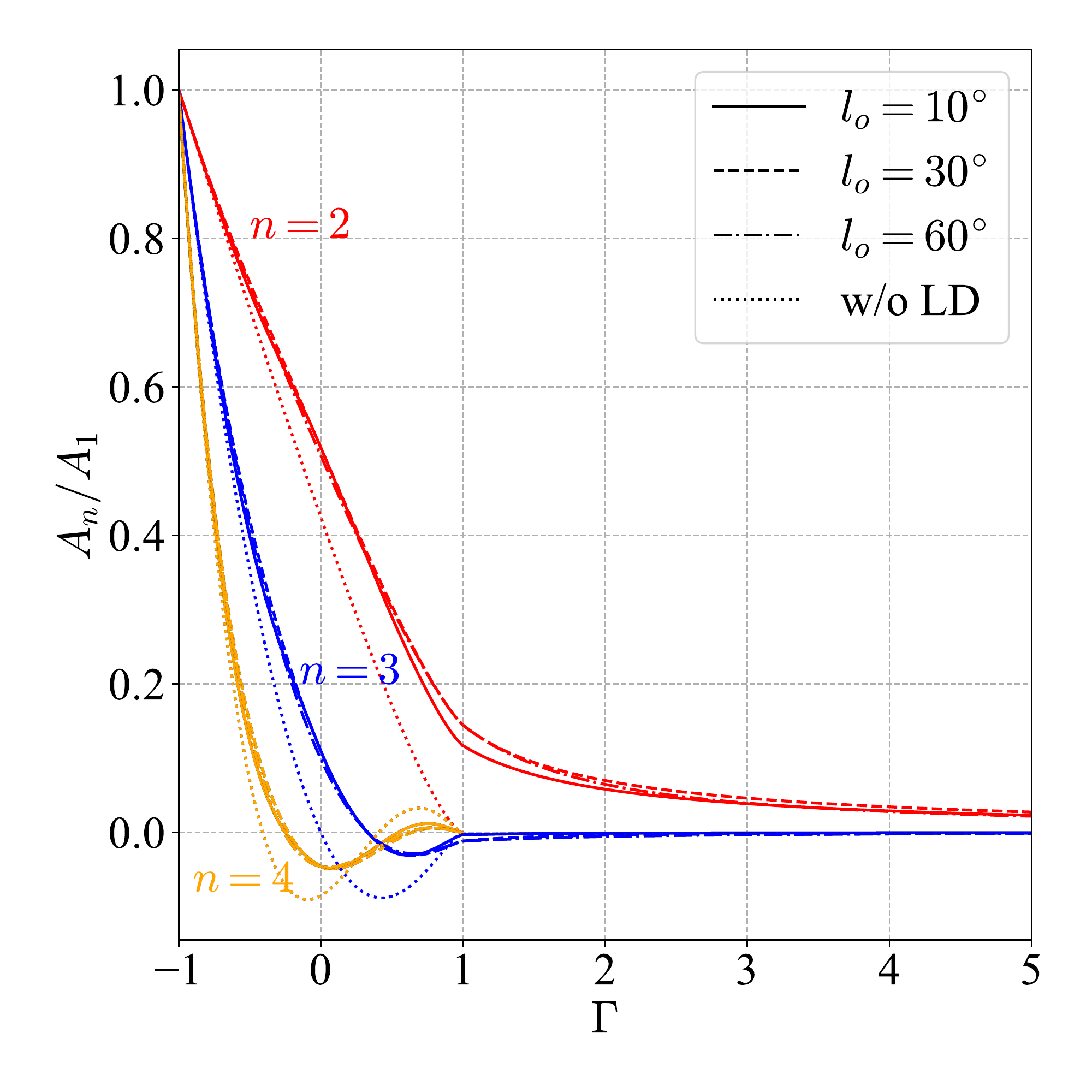} 
 \end{center}
 \caption{Ratios of Fourier coefficients of the photometric
     modulation due to a single spot, $A_n/A_1$, plotted against
     $\thetac$ ({\it left}) and $\Gamma$ ({\it right}).  Red, blue,
     and orange lines indicate $A_2/A_1$, $A_3/A_1$, and $A_4/A_1$,
     respectively.  We assume the limb darkening parameters of
     $\alpha_2=\alpha_{2\odot}$ and $\alpha_4=\alpha_{4\odot}$, and
     plot those ratios in solid ($\ello=10^\circ$), dashed
     ($\ello=30^\circ$), and dot-dashed ($\ello=60^\circ$) lines.  For
     reference, the results without limb darkening (w/o LD) are
     plotted in dotted lines.}
  \label{fig:a234}
\end{figure}

Figure \ref{fig:a234} plots the ratios of Fourier coefficients of the
single spot modulation, $A_n/A_1$ ($n=2, 3, 4$).  If limb darkening is
neglected, they reduce to $a_n/a_1$ that are a function of $\Gamma
\equiv \tan\ello\tan\ells$ (or $\thetac$) alone, which are
  plotted in dotted lines.  When the limb darkening effect is taken
  into account, $A_n/A_1$ depends on both $\ello$ and $\ells$.  As
  Figure \ref{fig:a234} implies, however, difference among the three
  curves for $\ello=10^\circ$, $30^\circ$, and $60^\circ$ is
  small. Thus, $A_n/A_1$ is still largely determined by the value of
  $\Gamma$ (or $\thetac$) even with limb darkening.
  
This result suggests that $A_n/A_1$ may be used to
  examine if the periodic signals detected from the observed
  photometric lightcurve are due to starspots, instead of other
  sources. It may be even possible to put a constraint on $\Gamma$
  from $A_n/A_1$ in principle. Since $\ello$ is equivalent to the
  stellar inclination for the observer that can be independently
  measured from either spectroscopy or asteroseismology
  \citep[e.g.,][]{Kamiaka2018,Kamiaka2019,SS2021}, the constraint on
  $\Gamma$ is translated to that on the spot latitude $\ells$.  In
  reality, it is feasible to derive a robust constraint on $\Gamma$
  only for a single spot case. The statistical distribution of
  $A_n/A_1$ for multi-spots is more useful to constrain the
  differential rotation as discussed below.

\section{Multiple starspots: model predictions and mock data analysis}

\subsection{Superposition of multiple starspots}

If more than one starspots are involved, we have to take into account
their relative phases, namely $\varphisinit$ in equation
(\ref{eq:varphis-t}), as well. In that case, equation
(\ref{eq:FTLt-spot-ld}) can be generalized to
\begin{eqnarray}
\label{eq:FTLt-spot-ld-2}
&& L_{\rm s}(t) = - \frac{\bs}{\pi} \left(1 - \frac{u_1}{3} -
\frac{u_2}{6}\right)^{-1} \Big[ \frac{A_0}{2} +
\sum_{n=1}^{\infty} A_n \cos n (\omegas t+\varphisinit) \Big] \cr
&& \quad=  - \frac{\bs}{\pi} \left(1 - \frac{u_1}{3} -
\frac{u_2}{6}\right)^{-1} \Big[ \frac{A_0}{2} +
\sum_{n=1}^{\infty} A_n \left(\cos n\varphisinit \cos n \omegas t
-\sin n\varphisinit \sin n \omegas t
\right) \Big] .
\end{eqnarray}

Thus, the lightcurve due to multispots becomes the superposition of
the following form:
\begin{eqnarray}
\label{eq:FTLt-multispot-ld}
 L_{\rm s}(t) && =  - \frac{1}{\pi} \left(1 - \frac{u_1}{3} 
-\frac{u_2}{6}\right)^{-1}\cr
&&  \times \Big\{\sum_{i=1}^{N_s} \bsi
\Big[ \frac{A_{0,i}}{2} +
\sum_{n=1}^{\infty} A_{n,i} \left(\cos n\varphisiniti
\cos n \omegasi t
-\sin n\varphisiniti \sin n \omegasi t
\right) \Big] \cr
&& ~ + \sum_{i=N_s+1}^{N_s+\tilde{N_s}} \bsi
\Big[ \frac{\tilde{A}_{0,i}}{2} +
\sum_{n=1}^{3} \tilde{A}_{n,i} \left(\cos n\varphisiniti
\cos n \omegasi t
-\sin n\varphisiniti \sin n \omegasi t
\right) \Big] \Big\} .
\end{eqnarray}
In the above equation, $N_s$ and $\tilde{N}_s$ denote the number of
spots with $|\Gamma|<1$ and $\Gamma>1$, respectively, $\bsi$ is the
amplitude of the photometric variation,
$\varphisiniti$ is the initial phase, $\omegasi=\omega(\ell_{{\rm
    s},i})$ is the angular frequency, and $A_{n,i}$ and
$\tilde{A}_{n,i}$ are the Fourier components, of the $i$-th starspot.

\subsection{Mock lightcurves and the Lomb-Scargle power spectra
\label{subsec:mock-LS}}

In order to examine to what extent one can extract the characteristic
signature of starspots from photometric stellar lightcurves, we create
mock lightcurves in the time domain, and compute the Lomb-Scargle
power spectra.  Our fiducial set of parameters is listed in Table
\ref{tab:parameter}.

The key parameter characterizing the spot in our model
  is $\bs$.  As described in subsection \ref{subsec:singlespot-wo-ld},
  $\bs$ is defined as $\As/\Rs^2$ in our model.
  \citet{2015ApJ...800...48M} found that the flux-weighted effective
  area, $\As$, for the Solar spot empirically obeys the
  Weibull distribution:
\begin{equation}
  \label{eq:weibull}
  f(\As; k, \lambda) d\As = k \left( \frac{\As}{\lambda} \right)^{k-1}
  e^{-(\As/\lambda)^k} \frac{d\As}{\lambda}
\end{equation}
from the observed photometric variation over years. The
  Weibull distribution is written in terms of $\As/\lambda$, the
  amplitude of the resulting spot modulation is simply scaled to the
  adopted value of $\lambda$. 

The expectation value of $\As$ from equation (\ref{eq:weibull}) is
\begin{equation}
  \label{eq:mean-weibull}
  \langle \As \rangle = \int_{A_{\rm th}}^\infty   \As f(\As; k, \lambda) d\As
  = \lambda \Gamma(1+\frac{1}{k}), 
\end{equation}
where $\Gamma(x)$ denotes the Gamma function and
$\Gamma(1+\frac{1}{k}) \approx 1.75$ for the solar value of $k=0.54$.
Also the corresponding cumulative number distribution of $\As$
exceeding the threshold value $A_{\rm th}$ is
\begin{equation}
  \label{eq:cumulative-weibull}
  P(\As>A_{\rm th}) = \int_{A_{\rm th}}^\infty f(\As; k, \lambda) d\As
  =e^{-(A_{\rm th}/\lambda)^k}.
\end{equation}
For instance, the top 10 percentile of spots have
$\As> 1.47\lambda$.

The best-fit values of the two parameters, $k$ and $\lambda$, vary for
different definitions of spots and different datasets
\citep{2015ApJ...800...48M}.  For definiteness, we adopt ``Sunspot
Umbral Area'' from the Helio-seismic and Magnetic Imager on the Solar
Dynamics Observatory (see their Table 1 ), and adopt $k=0.54$ and
$\lambda = 2.88 ~{\rm\mu Hem}=2.88\times10^{-6} (2\pi \Rs^2)$.  It is
likely that different stars may have different values of $k$ and
$\lambda$.  Since our model is fully analytical, however, it is
readily applicable for other choices. Thus, we fix their values below,
and generate mock data for multi-spots.
  
Equation (\ref{eq:mean-weibull}) suggests that a characteristic
  amplitude of the dimensionless parameter $\bs$ of our spots is
\begin{equation}
\label{eq:bspot-lambda}
\langle\bs\rangle \equiv \frac{\langle\As\rangle}{\Rs^2}
\approx \Gamma(1+1/k) \frac{\lambda}{\Rs^2} =
  2\pi \left(\frac{\lambda}{1 ~{\rm\mu Hem}}\right)\Gamma(1+1/k)~
  {\rm ppm},
\end{equation}
where the factor of $2\pi$ comes from the fact that $\lambda$ is
  given relative to the area of hemisphere (Hem), $2\pi \Rs^2$. Thus
we can safely neglect the finite size effect of an individual spot,
which is consistent with the assumptions of our analytic model, for
our adopted value of $\lambda =2.88~{\rm\mu Hem}$).

In order to understand the meaning of equation (\ref{eq:bspot-lambda}),
  let us define the effective radius of the spot $\rs$ through
\begin{equation}
\label{eq:def-rspot}
\As = \pi \rs^2.
\end{equation}
Substituting equation (\ref{eq:bspot-lambda}) into equation
(\ref{eq:def-rspot}), one obtains
\begin{equation}
\label{eq:rspot-Rs}
\frac{\rs}{\Rs} = \sqrt{\frac{\bs}{\pi}}
\approx 1.8\times 10^{-3}\sqrt{\frac{\bs}{10~{\rm ppm}}}
\approx 0.1^\circ \sqrt{\frac{\bs}{10~{\rm ppm}}},
\end{equation}
or equivalently
\begin{equation}
\label{eq:rspot-size}
\rs \approx 0.19R_\oplus \sqrt{\frac{\bs}{10~{\rm ppm}}}
\left(\frac{\Rs}{R_\odot}\right) .
\end{equation}
Equations (\ref{eq:rspot-Rs}) and (\ref{eq:rspot-size}) correspond to
the angular and real size corresponding to $\rs$ in terms of $\bs$.

We generate $N_{\rm tot}$ spots with $\bs$ following the Weibull
distribution, equation (\ref{eq:weibull}).  We adopted $N_{\rm
    tot}=30$ for definiteness so as to roughly reproduce the Solar
  spots.  The corresponding fraction of spots over the entire stellar
  surface may be computed from equation (\ref{eq:mean-weibull}):
\begin{equation}
\label{eq:area-fraction}
F(k,\lambda) = N_{\rm tot} \frac{\langle \As \rangle}{4\pi \Rs^2}
= 7.6\times10^{-5} \left(\frac{N_{\rm tot}}{30}\right)
\left(\frac{\lambda}{2.88~\mu{\rm Hem}}\right)
\left(\frac{\Gamma(1+1/k)}{1.75}\right).
\end{equation}
The value of $N_{\rm tot}$ is sensitive to the threshold value $A_{\rm
  th}$ in identifying a single spot even for the Sun, and moreover is
not clear for other stars. Our analytic formulation can be applied to
a different choice of $N_{\rm tot}$ in a straightforward manner.

The latitudes of spots $\ells$ are drawn from the isotropic
distribution function ($\propto |\sin\ells|$) but over the restricted
range of $-\ell_{\rm s,max}<\ells<\ell_{\rm s,max}$. We choose
  $\ell_{\rm s,max}=30^\circ$ as our fiducial value, but consider
  $75^\circ$ as well to examine its impact. The initial phases are
selected randomly for $0<\varphis<2\pi$.

For a given value of the observer's latitude $\ello$, we classify
  each spot according to $|\Gamma|<1$ and $\Gamma>1$, and compute the
  number of such spots $N_{\rm s}$ and $\tilde{N}_{\rm s} (=N_{\rm
    tot}-N_{\rm s})$, respectively.  Then the lightcurve modulation
  due to those spots is computed from equation
  (\ref{eq:FTLt-multispot-ld}).

We generate the mock lightcurves with cadence $T_{\rm samp}$ over the
duration of $T_{\rm obs}$.  We set the fiducial values as $T_{\rm
  samp}=30$ mins and $T_{\rm obs}=90$ days, following the long cadence
observation for one single quarter of the Kepler dataset.

Finally, we add the Gaussian noise to the lightcurves:
\begin{equation}
  f(x)= \frac{1}{\sqrt{2\pi\sigma_{\rm n}^2}}
  \exp\left(-\frac{x^2}{2\sigma_{\rm n}^2}\right).
\end{equation}
In what follows, we consider two cases, $\sigma_{\rm n}=0$ (noiseless)
and $\sigma_{\rm n}= 35 {\rm ppm}$ as a typical value for
  the Kepler data \citep[c.f.,][]{Walkowicz2013,Basri2020}, for
  simplicity.  Equation (\ref{eq:weibull}) implies that the flux
  modulation induced by a single spot is typically much smaller than
  the noise:
\begin{equation}
\label{eq:typical-amplitude}
\frac{\langle\bs\rangle}{\pi} = 10
\left(\frac{1.75 \lambda}{5.04 ~{\rm\mu Hem}}\right) {\rm ppm}.
\end{equation}
Thus, in the case of $\sigma_{\rm n}=35$ ppm, the clear periodic
signal is visible only for a relatively big spot ($\As>5\lambda$,
roughly corresponds to the top 10 percentile) or a clustered group of
nearby spots.

\begin{table}
  \tbl{Fiducial parameters for mock lightcurves}{%
  \begin{tabular}{llll}
symbol & range & note  \\ 
  \hline
  $\bs$ & $k=0.54$, $\lambda=2.88\mu$Hem
  & the Weibull distribution \\ 
$N_{\rm tot}$ & 30& total number of generated starspots\\
$\varphi_s$ & $[0, 2\pi]$ & uniform \\ 
  $\ells$ & $|\ells|<\ell_{\rm s,max}=30^\circ$
  & $P(\ells) \propto |\sin\ells|$ \\ 
$\ello$ & $0^\circ$
  & stellar inclination relative to the observer's line-of-sight \\ 
$\alpha_2$ & $\alpha_{2\odot}=0.163$  & differential rotation coefficient \\ 
$\alpha_4$ & $\alpha_{4\odot}=0.121$  & differential rotation coefficient \\ 
$u_1$ & 0.47  & linear limb-darkening parameter \\ 
$u_2$ & 0.23 & quadratic limb-darkening parameter \\ 
$2\pi/\omega_0$ & 10 days  & equatorial rotation period \\ 
$T_{\rm samp}$ & 30 mins & cadence of the observation \\
$T_{\rm obs}$ & 90 days & duration of the observation \\
\end{tabular}}\label{tab:parameter}
\begin{tabnote}
\end{tabnote}
\end{table}

\begin{figure}
 \begin{center}
 \includegraphics[width=16cm, bb = 0 0 829 550]{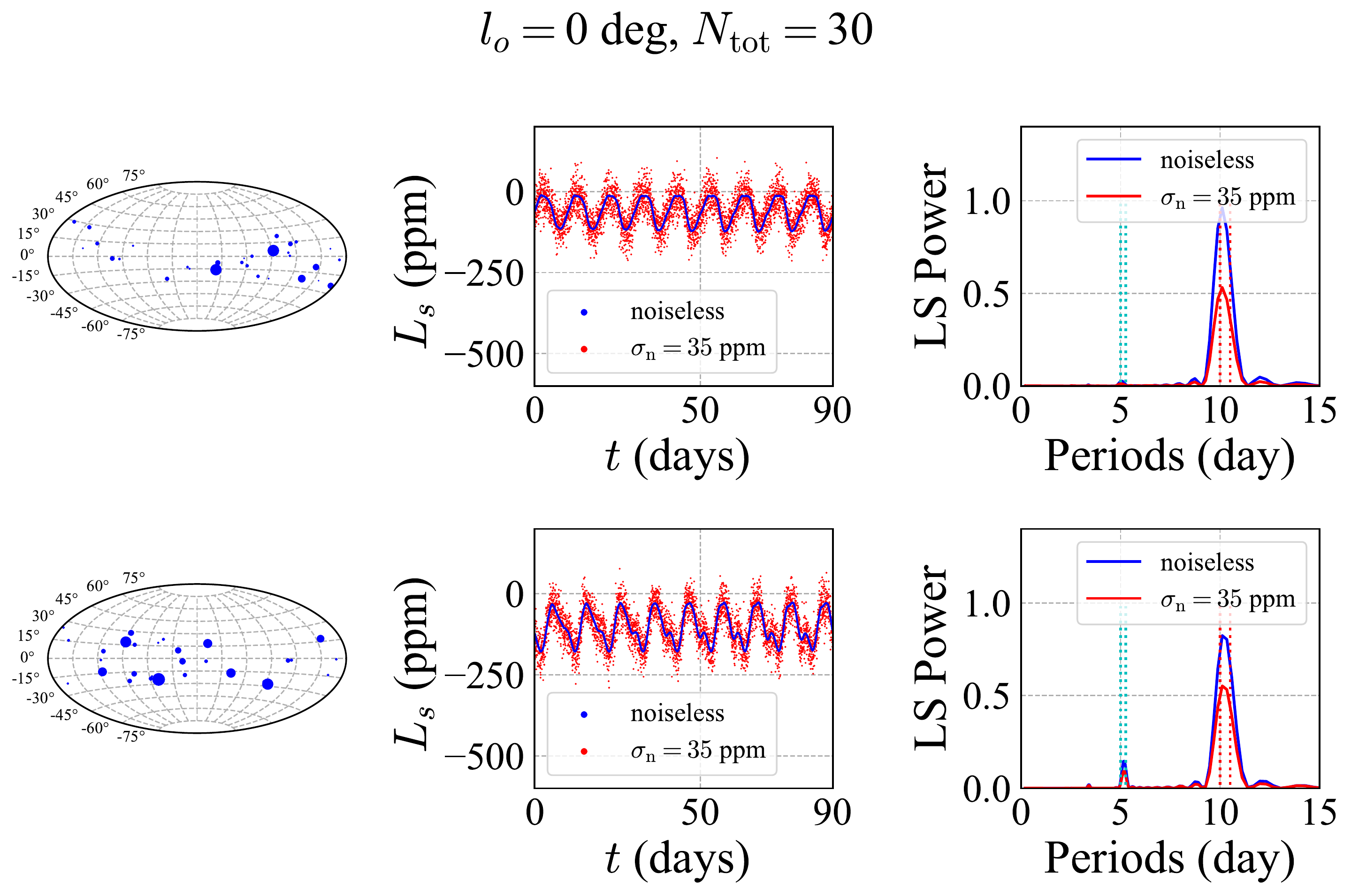}
 \end{center}
 \caption{Lightcurve modulations due to starspots ($N_{\rm tot}=30$, 
   $\ell_{\rm s,max}=30^\circ$) for an observer at $\ello=0^\circ$.
   Left, center, and right panels show the spot distribution on the
   stellar surface, normalized lightcurve, and the corresponding
   Lomb-Scargle power spectrum. Results for $\sigma_n=0$ and $35$ppm
   are plotted in blue and red, respectively. Upper and lower panels
   are different realizations of the statistically same model. The
   vertical dotted lines indicate the range of the differentially
   rotation periods; $P_{\rm spin}(\ells=0^\circ)/2$, $P_{\rm
     spin}(\ells=\ell_{\rm s,max})/2$, $P_{\rm spin}(\ells=0^\circ)$,
   and $P_{\rm spin}(\ells=\ell_{\rm s,max})$.}
 \label{fig:mock-lo0-Ntot30}
\end{figure}

Figure \ref{fig:mock-lo0-Ntot30} shows the mock data for two different
sets of realizations of starspots ($N_{\rm tot}=30$ $\ell_{\rm
  s,max}=30^\circ$) for an observer at $\ello=0^\circ$. The
  fractional area covered by spots varies from 0.3 to 1.3 times the
  expectation value of equation (\ref{eq:area-fraction}).

We search for periodic signals of an angular
frequency $\omega$:
\begin{eqnarray}
\label{eq:LS-nterms}
S(t) = S_0 +
\sum_{n=1}^{nterms} (S_{2n-1} \sin n \omega t + S_{2n} \cos n \omega t)
\end{eqnarray}
embedded in the mock lightcurves ({\it center panels}) using the
Lomb-Scargle (LS) method \citep{Lomb1976,Scargle1982}. The
conventional LS adopts $nterms=1$, and we compute the normalized
power spectrum:
\begin{eqnarray}
\label{eq:LSpower}
P(\omega) = 1- \frac{\chi^2(\omega)}{\chi_{\rm ref}^2},
\end{eqnarray}
where $\chi^2(\omega)$ is the residuals of the fit with $\chi_{\rm
  ref}^2$ being the reference value for a constant model.

Left panels of Figure \ref{fig:mock-lo0-Ntot30} plot the spot
distribution at $t=0$ for two different realizations. The area of
  each circle is plotted in proportion to $\bs$, and approximately
  represents the true ratio of the spot area and the entire stellar
  surface (but neglecting the distortion due to the projection onto
  the plane).  Since those spots span a range of latitudes, the
resulting lightcurves (center panels) are not exactly periodic in the
time domain due to the latitudinal surface differential rotation.
Right panels of Figure \ref{fig:mock-lo0-Ntot30} plot the
corresponding LS power.  The highest peaks around 10 days are located
over a range of rotation periods spanning $P_{\rm
  spin}(\ells=0^\circ)$ and $P_{\rm spin}(\ells=\ell_{\rm s,max})$ due
to the differential rotation. The secondary peaks around 5 days are
the second harmonics.  The ratios of those amplitudes carry important
information on $\ells$ and $\ello$, and will be discussed later
(subsection \ref{subsec:harmonics}) using the LS analysis with
$nterms=2$.

Our mock data completely neglect the dynamics of spots (their creation
and dissipation, and motion on the stellar surface) over the duration
of the observation $T_{\rm obs}=90$days, that corresponds to the
duration of a single quarter of the Kepler dataset. In order to
empirically evaluate the effects of the spot dynamic, we create 10
totally independent realizations drawn from the statistically same
spot distribution, and compute each LS power and the average over the
10 realizations. The latter may be interpreted as the average LS power
of the entire Kepler observing period, that is made of up to 10
quarters.

\begin{figure}
 \begin{center}
 \includegraphics[width=12cm, bb = 0 0 825 834]{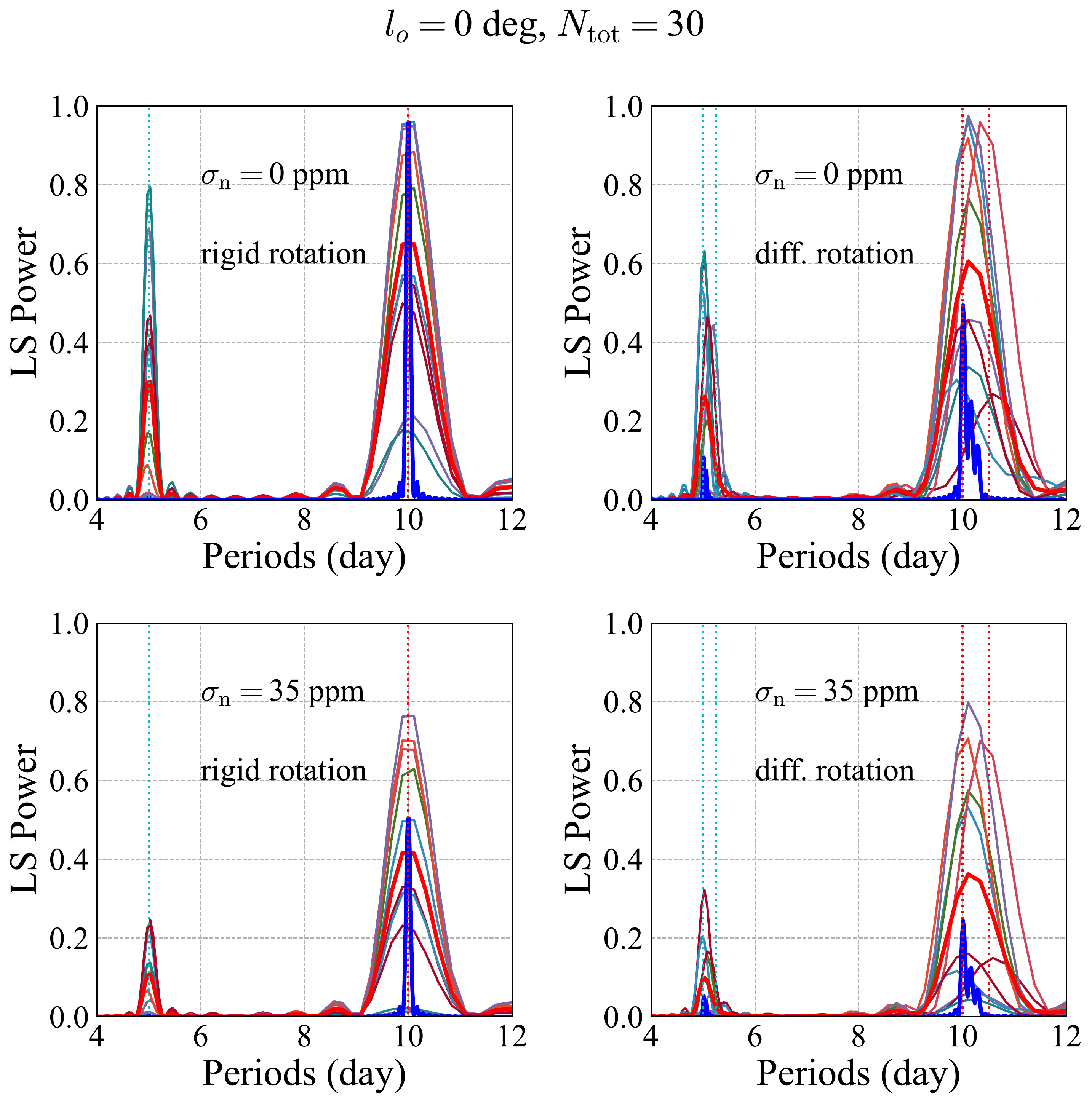} 
 \end{center}
 \caption{LS power spectra for different realizations corresponding to
   Figure \ref{fig:mock-lo0-Ntot30}. Upper and lower panels are for
   $\sigma_n=0$ and $35$ppm, and left and right panels are without
   and with differential rotation.  Thin curves (10 in total) indicate
   the results for different realizations of the spots for $T_{\rm
     samp}=30$ mins and $T_{\rm obs}=90$ days, with thick red lines
   being their average. Thick blue lines show the LS power for one
   realization but observed for $T_{\rm obs}=900$days. }
 \label{fig:LSpower-lo0-Ntot30}
\end{figure}

The results are shown in Figure \ref{fig:LSpower-lo0-Ntot30}.
Possible signatures of differential rotation may be found in the
variance among the LS power spectra for different quarters.  The width
of a peak with a detected period is determined by the entire duration
of the observation $T_{\rm obs}$, instead of the cadence $T_{\rm
  samp}$ in the present examples. For instance, one can resolve
  the periods for different spots only if they are static over $T_{\rm
    obs}=900$ days, but cannot for $T_{\rm obs}=90$ days.  While the
  non-evolving spots over $T_{\rm obs}=900$ days may not be so
  realistic in general, a small fraction of stars may have such
spots. Therefore our study suggests that it is worthwhile to attempt
searching for such signatures in the Kepler archive data.

\begin{figure}
 \begin{center}
 \includegraphics[width=16cm, bb = 0 0 829 549]{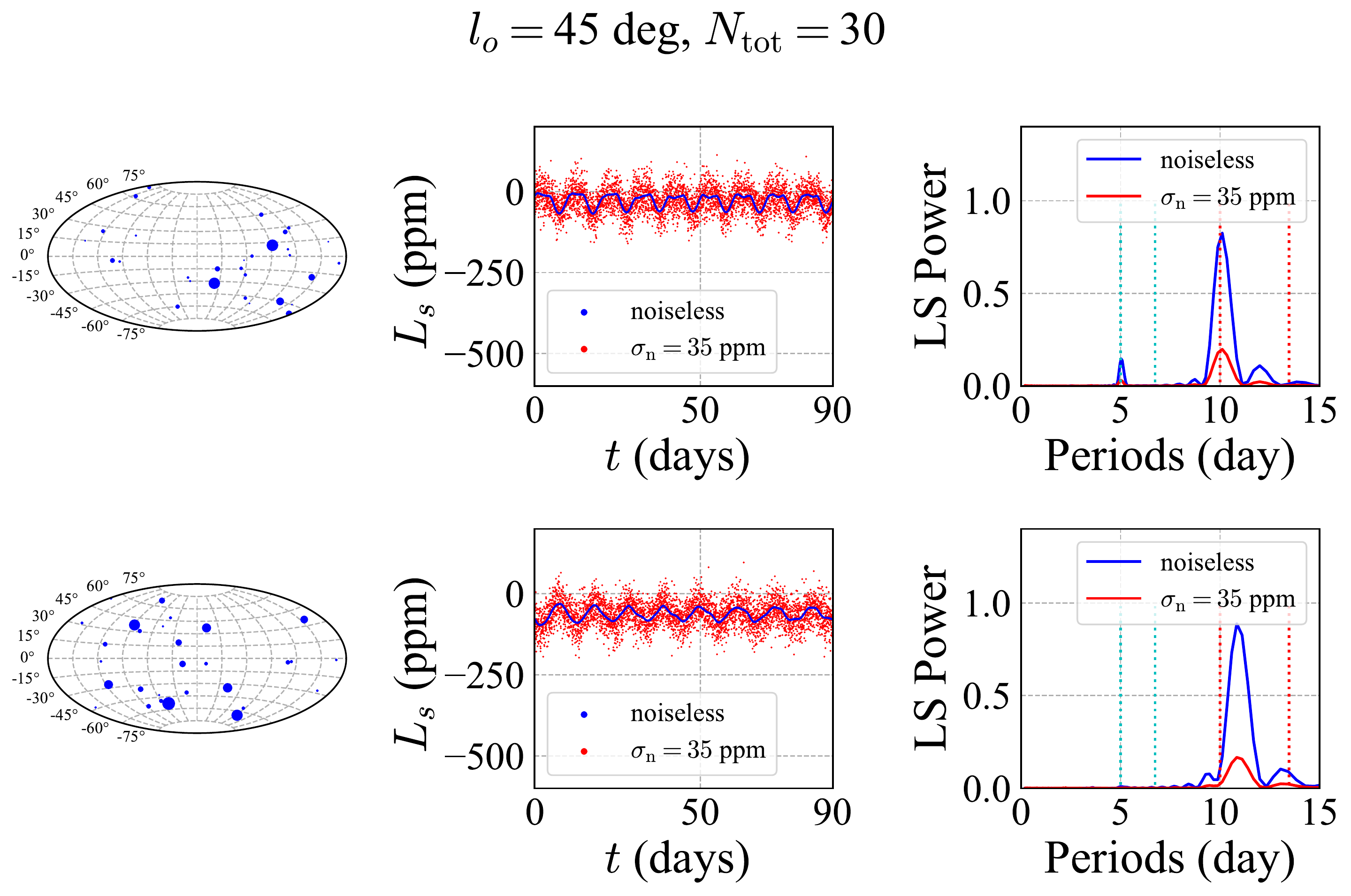} 
 \end{center}
 \caption{Same as Figure \ref{fig:mock-lo0-Ntot30}, but for
$\ell_{\rm s,max}=75^\circ$ and $\ello=45^\circ$. }
 \label{fig:mock-lo45-Ntot30}
\end{figure}

\begin{figure}
 \begin{center}
 \includegraphics[width=10cm, bb = 0 0 825 834]{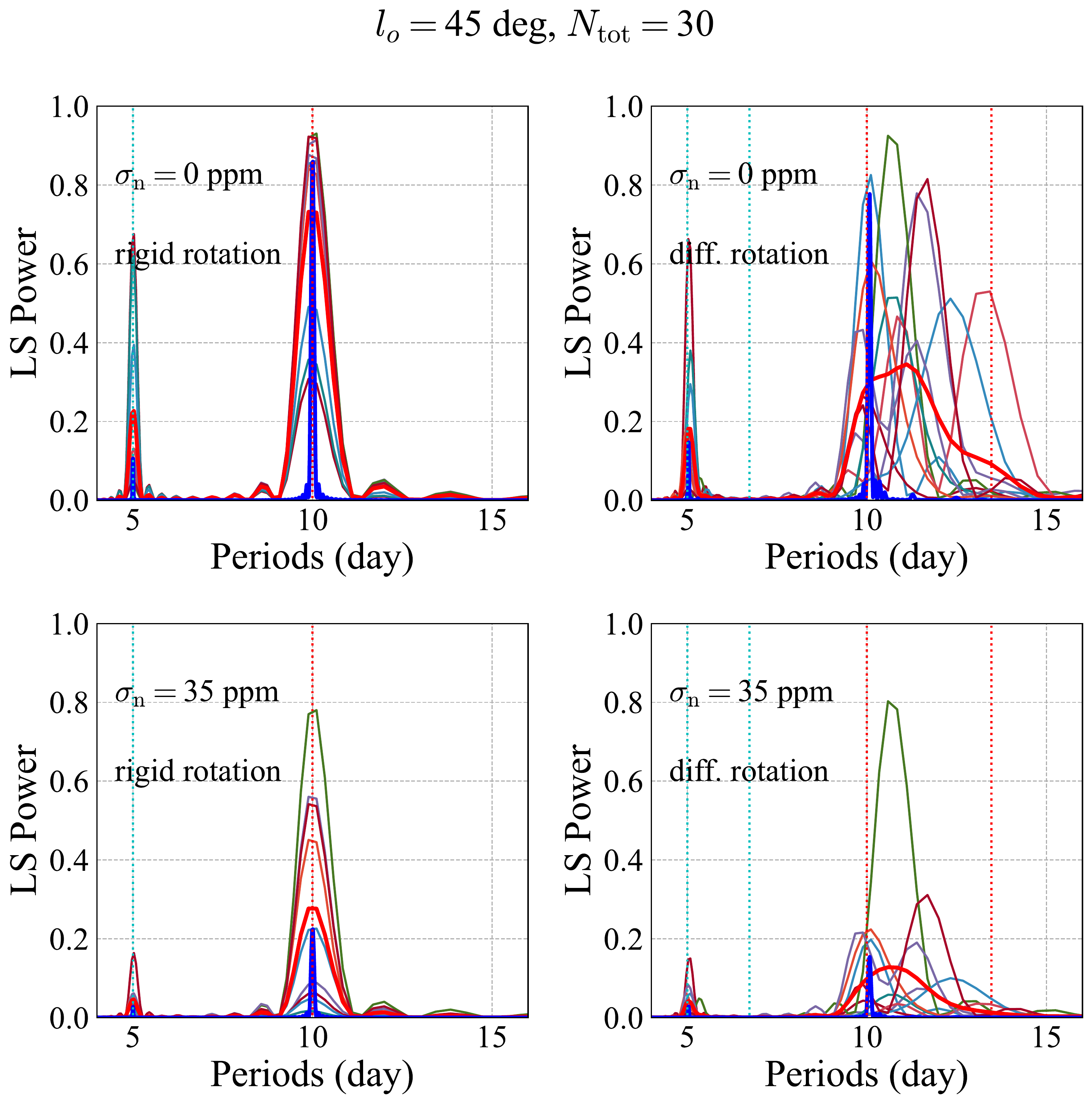}
 \end{center}
 \caption{LS power spectra for different realizations corresponding to
   Figure \ref{fig:mock-lo45-Ntot30}.}
 \label{fig:LSpower-lo45-Ntot30}
\end{figure}

Figures \ref{fig:mock-lo45-Ntot30} and \ref{fig:LSpower-lo45-Ntot30}
show the same plots as Figures \ref{fig:mock-lo0-Ntot30} and
\ref{fig:LSpower-lo0-Ntot30}, but for a wider distribution of spots
($\ell_{\rm s,max}=75^\circ$) observed from an observer located far
outside the stellar equatorial plane ($\ello=45^\circ$). As expected,
the effect of differential rotation is more visible than that for
$\ell_{\rm s,max}=30^\circ$ and $\ello=0^\circ$.

\begin{figure}
 \begin{center}
 \includegraphics[width=16cm, bb = 0 0 829 549]{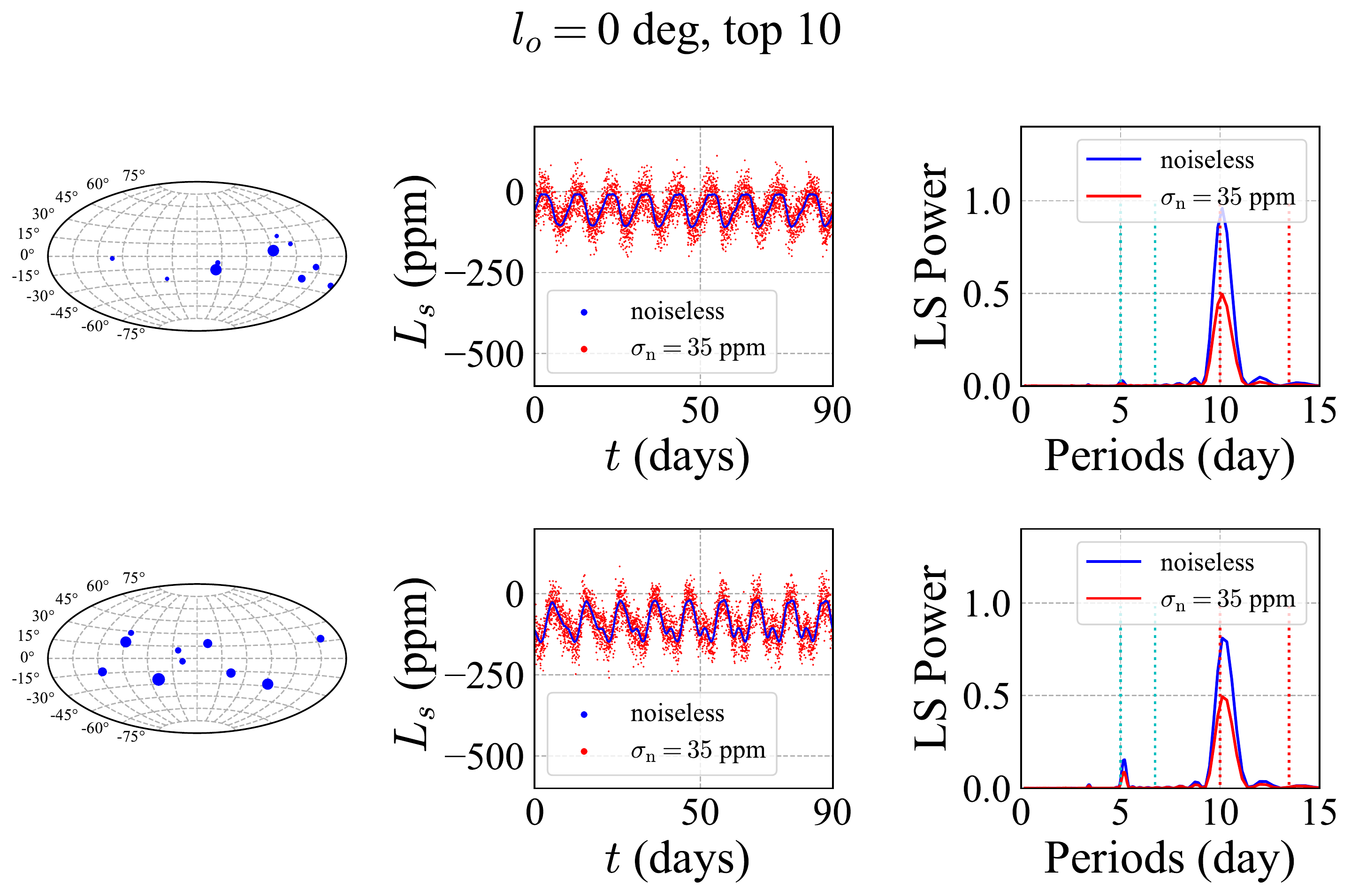} 
 \includegraphics[width=16cm, bb = 0 0 829 549]{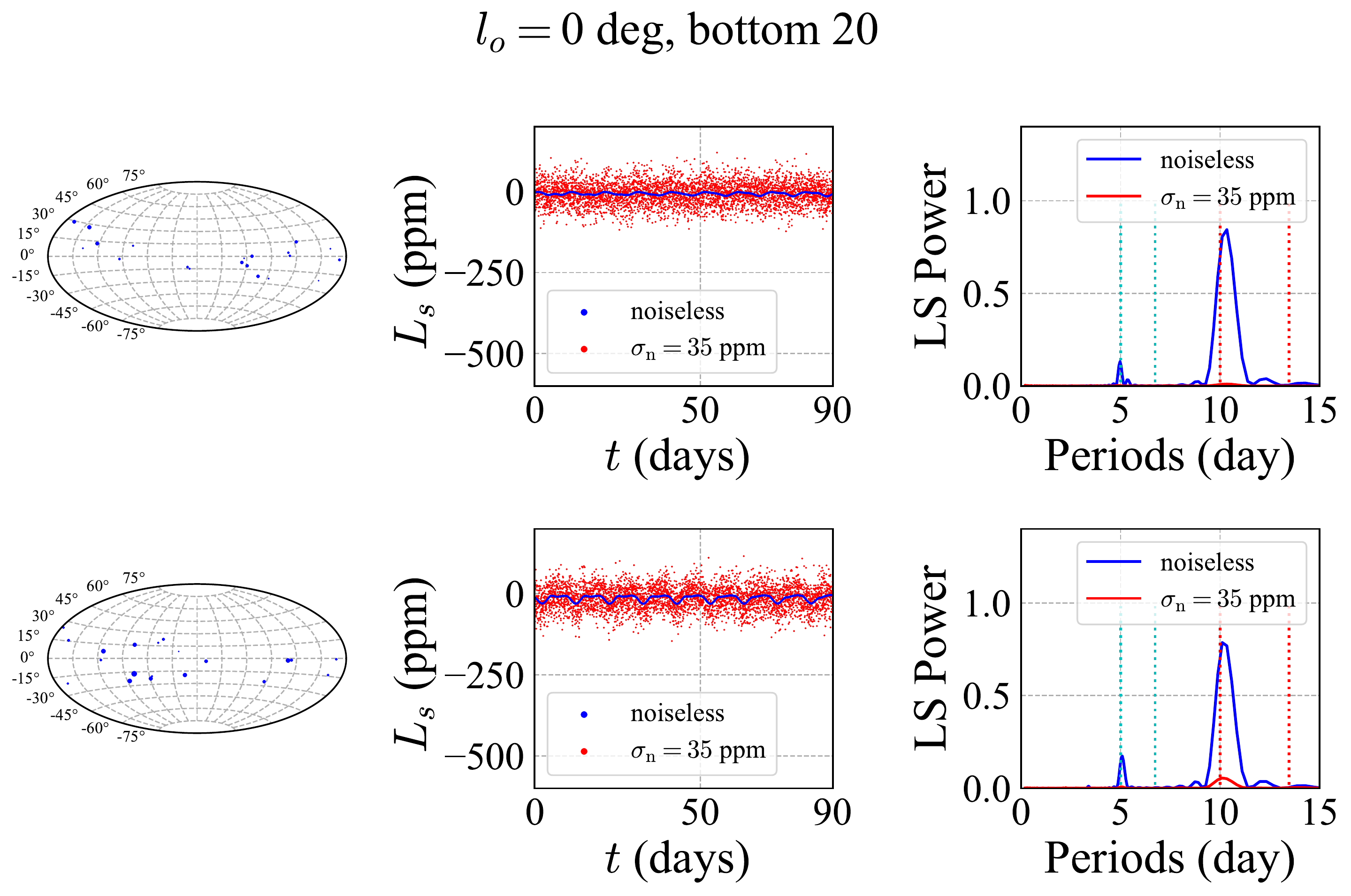} 
 \end{center}
 \caption{Same as Figure \ref{fig:mock-lo0-Ntot30},
   but computed for the top 10 spots ({\it upper two panels})
   and the remaining 20 spots ({\it lower two panels}).}
 \label{fig:mock-lo0-10-20}
\end{figure}

\begin{figure}
 \begin{center}
 \includegraphics[width=16cm, bb = 0 0 829 549]{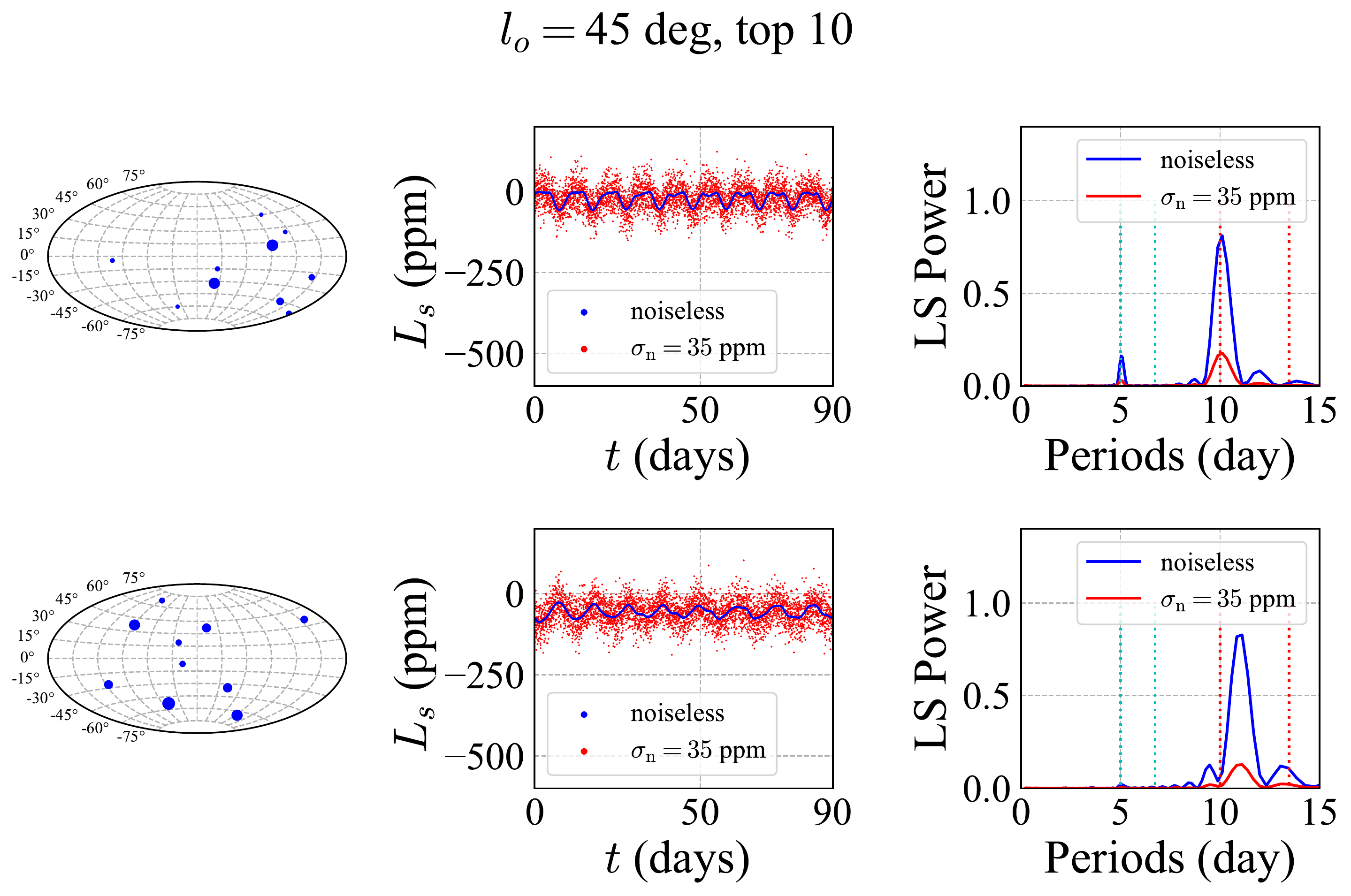} 
 \includegraphics[width=16cm, bb = 0 0 829 549]{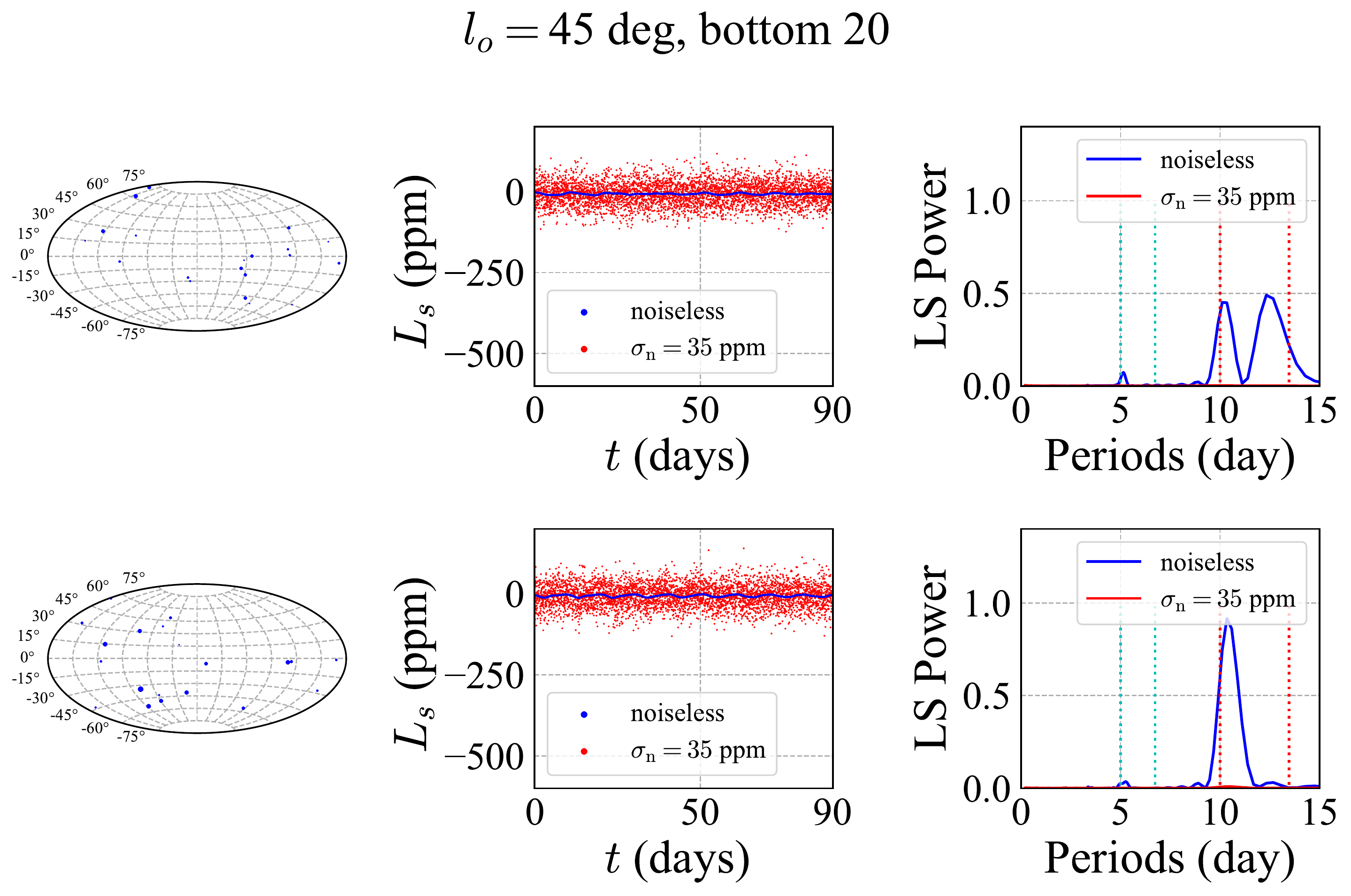} 
 \end{center}
 \caption{Same as Figure \ref{fig:mock-lo45-Ntot30},
   but computed for the top 10 spots ({\it upper two panels})
   and the remaining 20 spots ({\it lower two panels}).}
 \label{fig:mock-lo45-10-20}
\end{figure}

\begin{figure}
 \begin{center}
 \includegraphics[width=16cm, bb = 0 0 829 549]{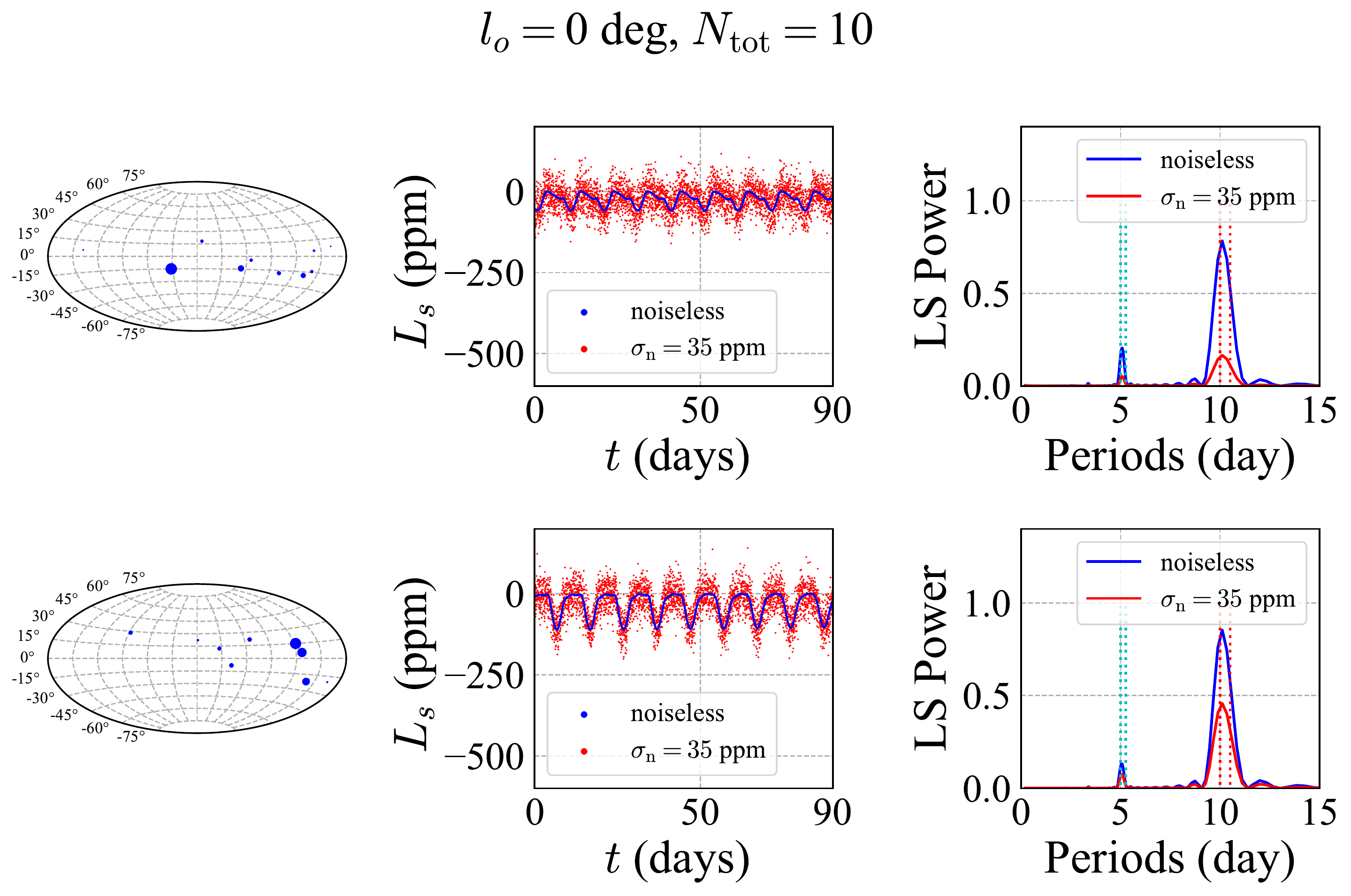} 
 \end{center}
 \caption{Same as Figure \ref{fig:mock-lo0-Ntot30}, but for
$N_{\rm tot}=10$. }
 \label{fig:mock-lo0-Ntot10}
\end{figure}

The visible periodicity of the lightcurve modulation in the center
panels of Figures \ref{fig:mock-lo0-Ntot30} and
  \ref{fig:mock-lo45-Ntot30} seems to be generated by a relatively
  small number of large spots.  To clarify this point, we repeated the
  analysis by dividing the 30 spots in the two realizations of Figures
  \ref{fig:mock-lo0-Ntot30} and \ref{fig:mock-lo45-Ntot30} separately
  into two groups; the top 10 spots and the remaining 20 spots.
  The resulting plots are shown in Figures \ref{fig:mock-lo0-10-20}
  and \ref{fig:mock-lo45-10-20}.  While those small spots still show
periodic signals in the {\it noiseless} lightcurve, they are
substantially buried in the case of our adopted noise of
$\sigma_n=35$ppm. In other words, the peaks in the LS power spectra
are dominated by a small fraction of spots, and should represent
mostly their properties (size, latitude, and rotation velocity), as
long as the Weibull distribution is a good approximation for the spot
distribution for stars other than the Sun. The above result also
  implies that our basic conclusion is not so sensitive to the choice
  of $N_{\rm tot}$; see Figure \ref{fig:mock-lo0-Ntot10}.

\subsection{Extracting the spot signature from the amplitude
  ratios of harmonics
\label{subsec:harmonics}}

We have shown that a single spot leaves a distinctive modulation
pattern in the Fourier coefficients of the harmonics (Figure
\ref{fig:a234}). In principle, the signature is important to
distinguish between the true and false rotation periods from
photometry. Nevertheless, it may be weakened for more realistic cases
of multispots, in particular under the presence of the stellar
differential rotation. We consider this question in detail using mock
data analysis.

\begin{figure}
 \begin{center}
 \includegraphics[width=7cm, bb = 0 0 536 536]{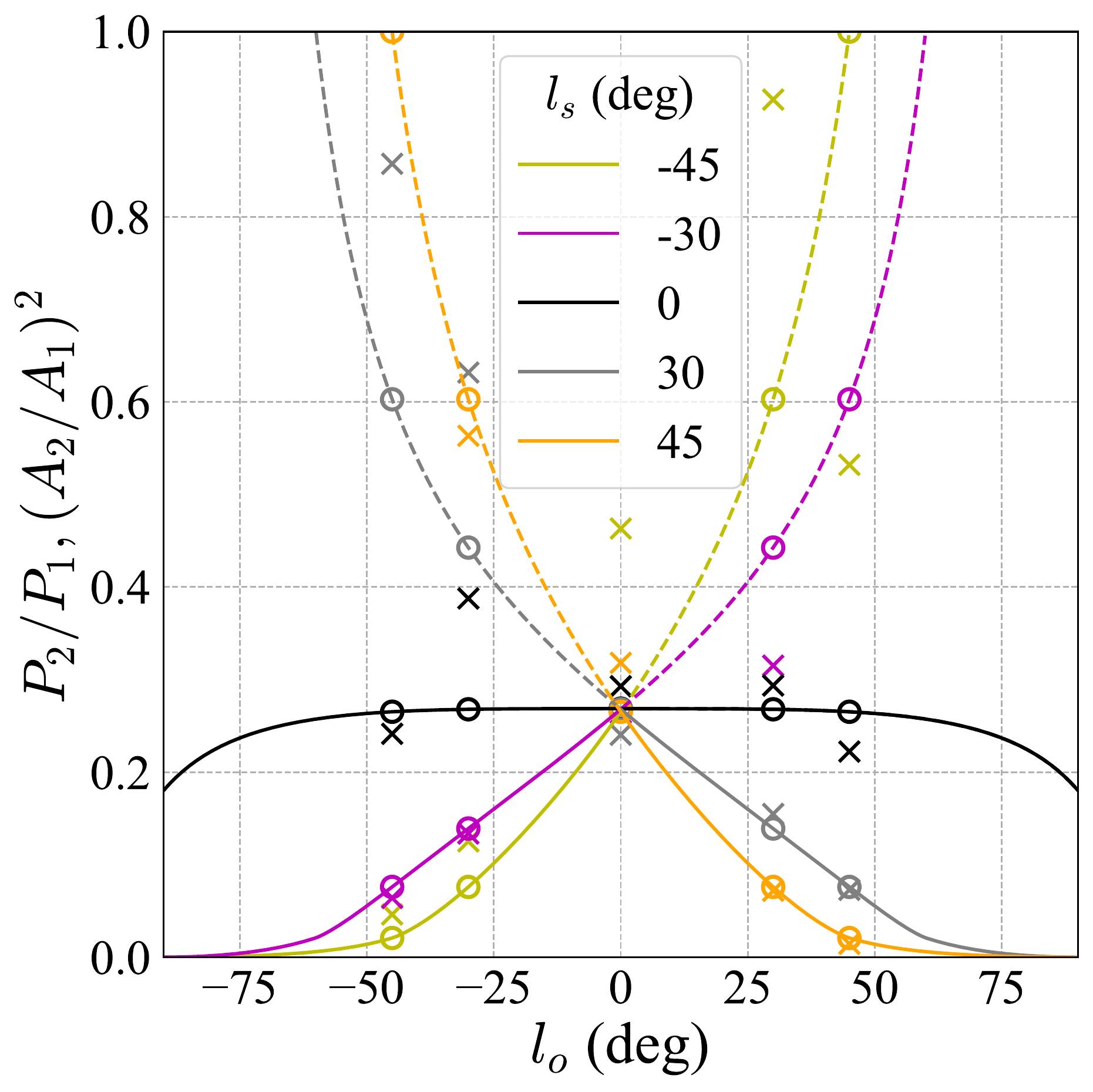}
 \includegraphics[width=7cm, bb = 0 0 536 536]{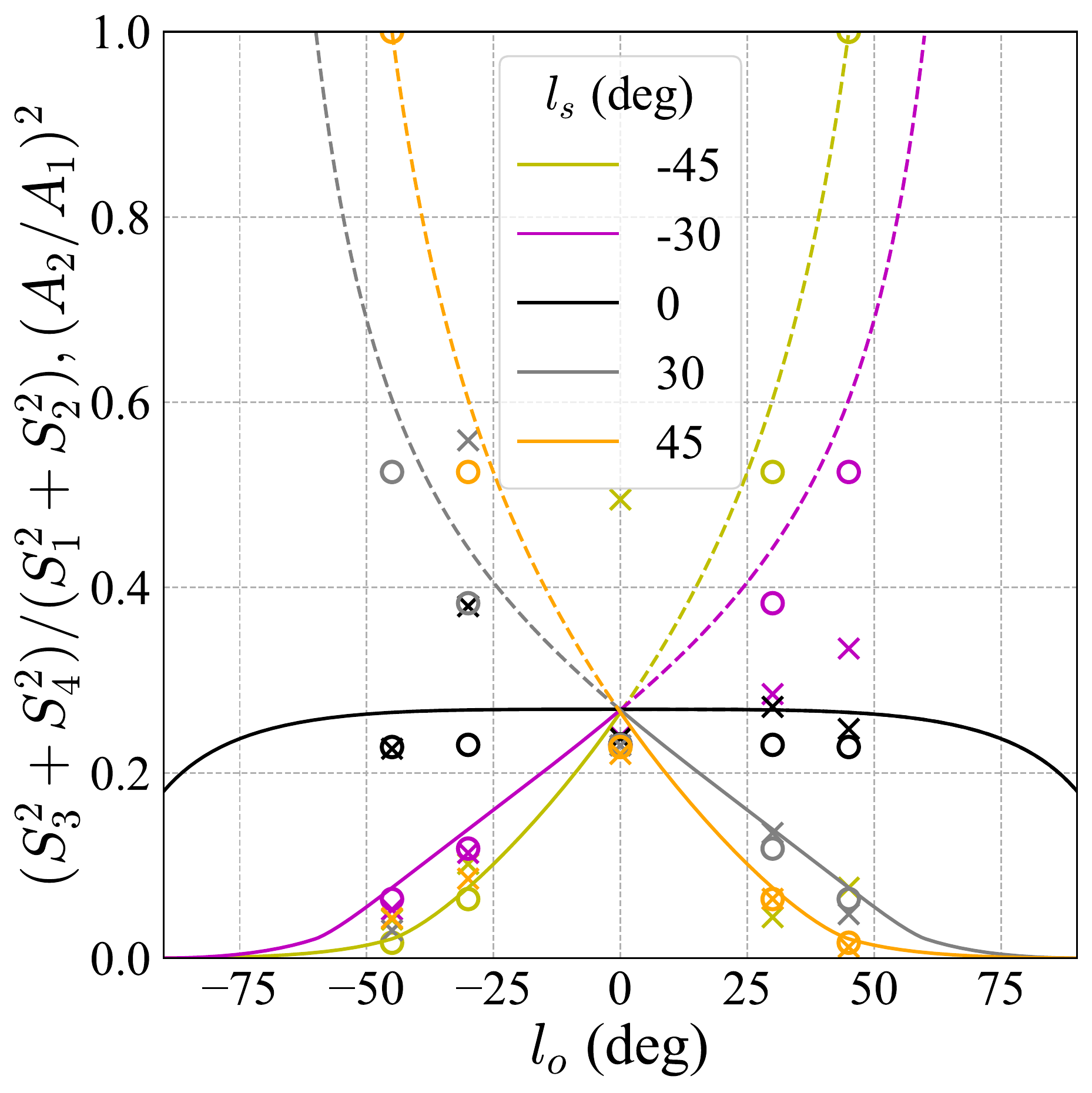}
 \end{center}
 \caption{Ratios of the second to fundamental Fourier coefficients for
   a single spot.  Theoretical predictions $(A_2/A_1)^2$ for
   $\ells=-45^\circ$ (green yellow), $-30^\circ$ (purple), $0^\circ$
   (green), $30^\circ$ (gray), and $45^\circ$ (orange) are plotted
   against $\ello$ in solid ($\ells \ello\ge 0$) and dashed ($\ells
   \ello<0$) lines.  Open circles and crosses indicate the ratios
   estimated from the mock lightcurves for $\bs\approx 85$ ppm with
   $\sigma=0$, and $35$ ppm, respectively.  Left and right panels plot
   the result based on the Fourier power spectrum and the LS analysis
   with $nterms=2$.}
 \label{fig:harmonic-ratio-single}
\end{figure}

Consider a single spot case.  Figure \ref{fig:harmonic-ratio-single}
compares the theoretical model predictions $(A_2/A_1)^2$ (solid lines)
for the stellar rotation period against the measurement from the mock
data for a single spot located at a given $\ells$ viewed from a
line-of-sight direction of $\ello$. We choose the value of $\bs
\approx 85$ ppm so that it corresponds to the top 10 percent of the
whole spot distribution, {\it i.e.,} $\As=4.7\lambda$ from equations
(\ref{eq:bspot-lambda}) and (\ref{eq:cumulative-weibull}).
Incidentally, the theoretical curve is invariant with respect to the
transformation of $(\ello, \ells) \leftrightarrow (\ells, \ello)$ as
equation (\ref{eq:Lt-spot-visible}) indicates.

We adopt two different estimators. One is based on the standard
Fourier power spectrum, and plots the corresponding amplitude ratio
$P_2/P_1$ (left panel).  The other is based on the LS analysis.  In
this case, we first identify the best-fit angular frequency
$\omega_{\rm fit}$ from the LS power spectra using $nterms=1$ in
equation (\ref{eq:LS-nterms}). Then we fit the data to equation
(\ref{eq:LS-nterms}) with $nterms=2$ by setting $\omega=\omega_{\rm
  fit}$, and obtain the Fourier coefficients $S_1$, $S_2$, $S_3$, and
$S_4$ simultaneously. The symbols in the right panel plot the ratio
$(S_3^2+S_4^2)/(S_1^2+S_2^2)$.

In the noiseless case, the Fourier power spectrum recovers the
theoretical predictions very well, but the LS analysis seems to
slightly but systematically underestimate the theoretical values.
We do not understand why, but the fit to equation
  (\ref{eq:LS-nterms}) with $nterms=2$ might be too restrictive and
  thus very sensitive to the best-fit value of $\omega_{\rm fit}$
  estimated from that with $nterms=1$.

In any case, the ratios estimated for data with $\sigma_{\rm n}=35$
ppm are not so accurate especially when the latitude of the spot is
significantly different from the observer's line-of-sight (with
different signs of $\ells$ and $\ello$, for instance). Therefore,
Figure \ref{fig:harmonic-ratio-single} implies that it is possible to
constrain $\ells$ and $\ello$ from the harmonic amplitude ratio for a
single spot at least for $\sigma_{\rm n}=35$ ppm.  For multi-spots
cases, however, we find that the amplitude ratio varies significantly
due to the differential rotation.  Thus this methodology seems to be
useful to constrain the spot parameter only when the photometric
signal is dominated by a single prominent region.

\subsection{Photometric rotation period for differentially rotating stars
\label{subsec:Prot}}

Time-dependent distribution of multi-spots over a stellar
  surface leads to complex photometric modulation signals.  Combined
  with the effect of latitudinal differential rotation, the peak of
  the rotation period would vary at different observing epochs. In
  turn, the variation of the rotation period among different quarters
  may constrain the degree of the differential rotation.

\begin{figure}
 \begin{center}
 \includegraphics[width=15cm, bb = 0 0 741 744]{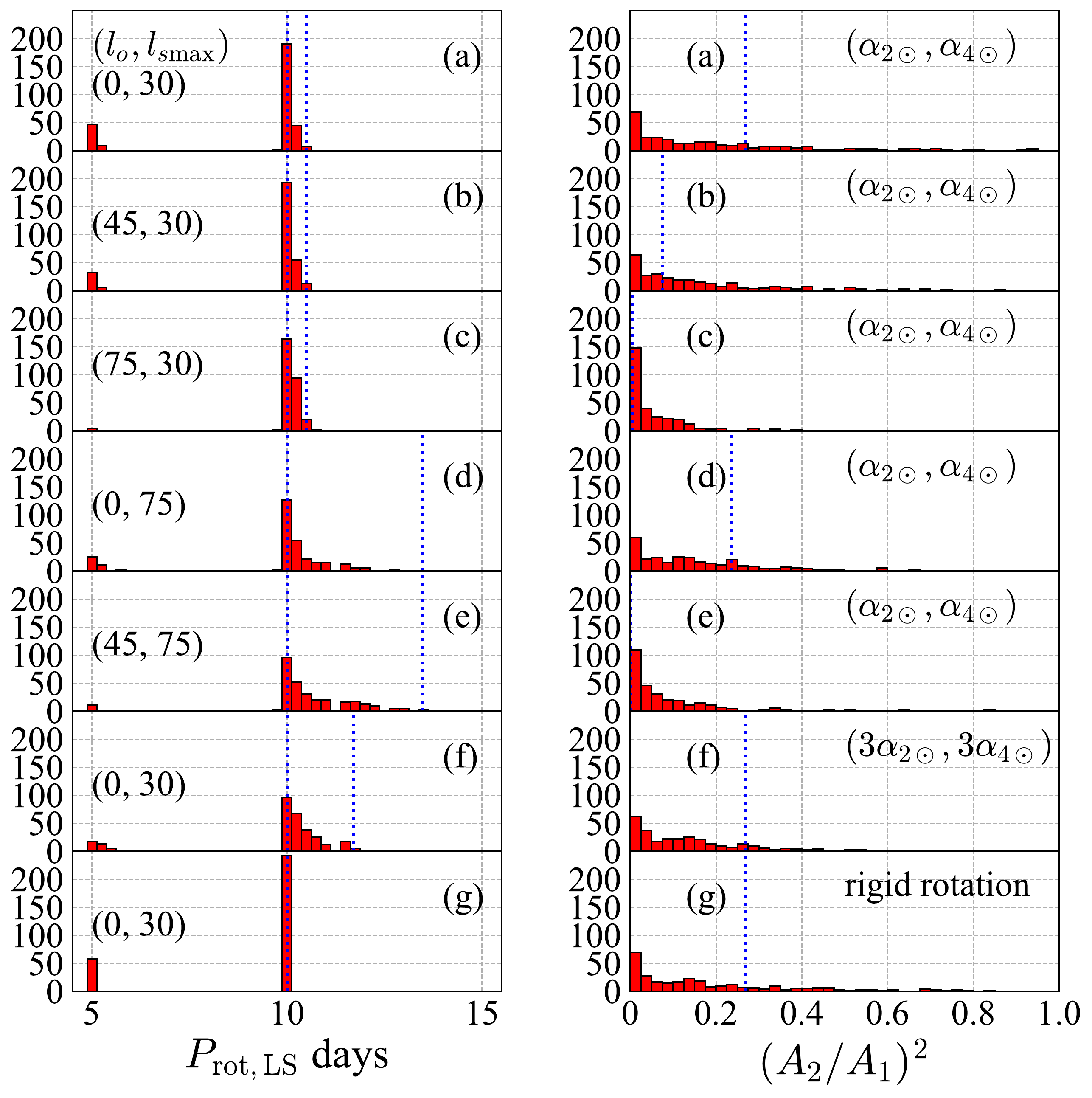}
 \end{center}
 \caption{Distribution of the photometric rotation period and the
   Fourier coefficient ratio estimated from 300 realizations of
   $N_{\rm tot}=30$ spots with $\sigma_{\rm n}=35$ppm. We adopt the
   fiducial parameter set of the equatorial rotation period $10$ days,
   $T_{\rm samp}=30$ mins, $T_{\rm obs}=90$ days. We compute the
   histograms by varying the observer's line-of-sight, and the range
   of the spot latitudes $-\ell_{\rm s,max}<\ells<\ell_{\rm s,max}$,
   and the differential rotation parameters, which are indicated in
   each panel.  Vertical dotted lines indicate the range of the
   differentially rotation periods as in Figure
   \ref{fig:mock-lo0-Ntot30} (left panels), and the predicted ratio
   for $\ell_{\rm s,max}$ (right panels). Panels (a), (b) and (c)
     use the same set of 300 realizations but viewed from the
     different observer's latitudes.}
   \label{fig:Prot-distribution}
\end{figure}

  In order to examine to what extent such signatures are indeed
  detectable from the Kepler data, we perform the LS analysis for
  seven different sets of mock data and plot distribution of the peak
  rotation period $P_{\rm rot, LS}$ in Figure
  \ref{fig:Prot-distribution}. Basically we adopt the fiducial values
  for parameters in Table \ref{tab:parameter}; the equatorial rotation
  period of 10 days, $N_{\rm tot}=30$ spots following the Weibull
  distribution with $k=0.54$ and $\lambda=2.88 \mu$Hem, the cadence of
  $T_{\rm samp}=30$ mins over an observing period of $T_{\rm obs}=90$
  days corresponding to one quarter of the Kepler long-cadence data.
  The latitude of the observer's line-of-sight $\ello$ and the range of the
  spot latitude $\ell_{\rm s,max}$ are indicated in each panel.

  The left panel of Figure \ref{fig:Prot-distribution} shows the
    histograms of the identified rotation period $P_{\rm rot, LS}$,
    while the right panel plots histograms of the corresponding
    harmonic amplitude ratio $(A_2/A_1)^2$.  Each histogram for seven
    models is computed from 300 realizations.  The first three panels
    (a), (b) and (c) assume the differential rotation law and spot
    pattern similar to the Sun.  They use the same 300 realizations of
    the spot pattern over $ -30^\circ \leq \ells \leq 30^\circ$, but
    viewed from $\ello=0^\circ$, $45^\circ$, and $75^\circ$,
    respectively.  Similarly, panels (d) and (e) share the same set of
    300 realizations with $-75^\circ \leq \ells \leq 75^\circ$ but
    viewed from $\ello=0^\circ$ and $45^\circ$, respectively.

  According to equation (\ref{eq:diff-law}), the rotation period of
  the surface is longer than its equatorial value (10 days), and the
  width of the distribution reflects the observed range of the spot
  latitudes $\ells$ and the values of $\alpha_2$ and $\alpha_4$.

  Difference among panels (a), (b), and (c) is simply due to the fact
  that the observer at higher $\ello$ preferentially sees the spots
  located at higher $\ells$ as clearly illustrated in Figures
  \ref{fig:ello0}, \ref{fig:ello30}, and \ref{fig:ello60}. Since the
  rotation periods estimated by observers at high $\ello$ should be
  dominated by a small number of big spots around $\ells > 0^\circ$,
  their distribution is shifted towards the larger $P_{\rm rot, LS}$
  due to the differential rotation, and the corresponding amplitude
  ratio becomes smaller as qualitatively expected from Figure
  \ref{fig:harmonic-ratio-single}.  A fraction of spot patterns may
  exhibit an approximate symmetry between $\varphis$ and
  $\varphis+\pi$ by chance, which would be interpreted as $P_{\rm rot,
    LS}=5$ days. Such symmetric patterns are more likely to be visible
  from the edge-on view ($\ello=0^\circ$), which explains the fraction
  of the second peak around $P_{\rm rot, LS}=5$ days in panels (a),
  (b) and (c).

  The next two panels (d) and (e) consider the case for the broader
  spot distribution over $-75^\circ \leq \ells \leq 75^\circ$. Because
  of the presence of a few spots located at higher latitudes, the
  differential rotation becomes more important, and the distribution
  of $P_{\rm rot, LS}$ becomes even broader towards its larger value.

 The last two panels are shown just for comparison purpose; panel (f)
 is for the stronger differential rotation case
 ($\alpha_2=3\alpha_{2\odot}$ and $\alpha_4=3\alpha_{4\odot}$), and
 panel (g) is for rigid rotation. Given the same spot distribution
 pattern, comparison among panels (a), (f) and (g) indicates how the
 differential rotation law affects the distribution of the rotation
 period at different quarters of the Kepler data, for instance.  This
 is expected to be directly applicable to put statistical constraints
 on the degree of latitudinal differential rotation of a population of
 stars, or to estimate the parameter $\alpha_2$ (and even $\alpha_4$)
 for stars exhibiting clear photometric lightcurve modulations. 

While the harmonic ratios shown in the right panels reflect the
statistical distribution of the spot latitudes to some extent, they
are sensitive to the spot area distribution and do not seem to provide
quantitatively useful information. Nevertheless, the histograms are
qualitatively consistent with the expected range of the ratios
plotted as the vertical dotted lines. 

\section{Summary and conclusion}

We have presented an analytic model of the lightcurve variation due to
starspots on a differentially rotating surface. If the dynamics of the
spots over the timescale of the observing period is neglected,
the Fourier coefficients of the harmonics of the rotation period are
written primarily in terms of the latitude of spots and the observer's
line-of-sight direction angle.

In order to understand the
  resulting lightcurve variations, we generate various realizations of
  starspots according to the analytic model, and
  compute the Lomb-Scargle power spectra for the mock datasets.

Even though our analytical model neglects the evolution of spots on
the stellar surface (dynamical motion, creation and annihilation), its
prediction provides a useful framework to interpret the photometric
variation of stars, in particular from the existing Kepler data and
the future space-born mission. The conclusion and implications of the
paper are summarized below.

 1) If a photometric lightcurve of a star exhibits a clear single
  peak in the LS periodogram, the star may be well approximated as a
  rigid rotator, and the peak should correspond to the rotation
  period.
  
  2) For those stars that have multiple peaks in the LS periodogram,
  the distribution of the peaks estimated in different quarters may be
  used to put constraints on parameters characterizing the
  differential rotation law.

  3) In principle, the ratio of harmonics for the rotation period may
  constrain the spot latitude $\ells$ and stellar inclination $\ello$
  given a limb darkening law. The constraint, however, is sensitive to
  the spot distribution, and seems to be useful only for a single spot
  dominated case. Nevertheless, joint analysis with independent
  constraints on $\ello$ from spectroscopic and/or asteroseismic
  measurements may improve the constraint.  We have not explored this
  possibility in the present paper, but it is worthwhile to pursue in
  future.
  
  4) The analytical model presented in the paper is based on the
    distribution of the {\it effective} area of spots $\bs$ alone, and
    does not require the information of the geometric area and
    temperature simultaneously. Thus it is applicable not only for
    spots on main-sequence stars, but for other inhomogeneities on
    rotating systems.  For instance, the recent discovery of
  the fastest-period white dwarf \citep{Kilic2021} indicates that the
  interpretation of the photometric modulation of white dwarfs is
  crucial in extracting their rotation period.  Since it is likely to
  originate from the small hot spot around the polar region, the
  determination of $\ells$ and $\ello$ with respect to our
  line-of-sight may be more promising for white dwarfs than for stars
  with many different spots, as long as the modulation signal-to-noise
  ratio is sufficiently high.

  The above findings may have numerous useful applications even in the
  existing Kepler data that cover a wide variety of stars with
  different properties of spots on their surface.  We are currently
  working on the joint analysis of photometric and asteroseismic
  measurements of Kepler stars selected by \citet{Kamiaka2018}, and
  plan to present the results elsewhere in due course (Y.Lu et al. in
  preparation).  

\bigskip

\section*{Acknowledgements}
We thank an anonymous referee for various constructive comments on
  the manuscript.  Simulations and analyses in this paper made use of
a community-developed core Python package for Astronomy, {\tt
  Astropy}.  This work is supported by Grants-in Aid for Scientific
Research by the Japan Society for Promotion of Science (JSPS)
No.18H012 and No.19H01947, and from JSPS Core-to-core Program
``International Network of Planetary Sciences''.

\bibliographystyle{apj}
\bibliography{ref-suto}

\appendix

\section{Fourier series expansion for spots
  with $|\Gamma|<1$ \label{sec:FT-ld}}

The present paper is based on the analytic Fourier series
  expansion of the photometric lightcurve due to a single spot.
  Those expressions are derived in this appendix.

We first compute the Fourier expansion of equation (\ref{eq:Lt-spot}) by
setting
\begin{equation}
\label{eq:Lt-spot-1}
F_1(t) \equiv \max(\cos\omegas t + \Gamma, 0)
= \frac{a_0}{2} + \sum_{n=1}^\infty a_n \cos n\omegas t.
\end{equation}
Thus, the coefficients $a_n$ are simply given by
\begin{equation}
\label{eq:an-1}
  a_n = \frac{\omegas}{\pi} \int_{-\pi/\omegas}^{+\pi/\omegas}
  \max(\cos\omegas t + \Gamma, 0) \cos n\omegas t dt.
\end{equation}
It is convenient to introduce the angle $\thetac$ for $|\Gamma|<1$
through
\begin{equation}
\Gamma \equiv -\cos\thetac \qquad (0<\thetac<\pi).
\end{equation}
Then, equation (\ref{eq:an-1}) reduces to
\begin{eqnarray}
\label{eq:an-2}
  a_n &=& \frac{2}{\pi} \int_{0}^{\thetac}
  (\cos\theta -\cos \thetac) \cos n\theta d\theta \cr
  &=&\frac{1}{\pi}
  \int_{0}^{\thetac} [\cos(n+1)\theta + \cos(n-1)\theta]d\theta
  -\frac{2}{\pi} \cos\thetac  \int_{0}^{\thetac} \cos n\theta d\theta.
\end{eqnarray}
The straightforward integration of equation (\ref{eq:an-2}) yields
\begin{eqnarray}
\label{eq:a0}
\frac{a_0}{2} &=& \frac{1}{\pi}
\left(\sin\thetac-\thetac\cos\thetac\right), \\
\label{eq:a1}
a_1 &=& \frac{1}{2\pi}\left(2\thetac -\sin2\thetac\right), \\
\label{eq:a2}
a_n &=& \frac{1}{n\pi}\Big[
\frac{\sin (n-1)\thetac}{n-1}
-\frac{\sin (n+1)\thetac}{n+1}\Big] \qquad (n \geq 2),
\end{eqnarray}
which are a set of coefficients shown in equation
(\ref{eq:Lt-spot-Fourier}) of the main text.

If the limb darkening effect is considered, one has to compute
  two additional expansions including
\begin{equation}
\label{eq:Lt-spot-2}
F_2(t) \equiv \left[\max(\cos\omegas t + \Gamma, 0)\right]^2
= \frac{b_0}{2} + \sum_{n=1}^\infty b_n \cos n\omegas t,
\end{equation}
and
\begin{equation}
\label{eq:Lt-spot-3}
F_3(t) \equiv \left[\max(\cos\omegas t + \Gamma, 0)\right]^3
= \frac{c_0}{2} + \sum_{n=1}^\infty c_n \cos n\omegas t.
\end{equation}
Similarly to equation (\ref{eq:an-2}), the coefficients $b_n$ and
$c_n$ are given as
\begin{eqnarray}
\label{eq:bn-1}
  b_n &=& \frac{2}{\pi} \int_{0}^{\thetac}
  (\cos\theta -\cos \thetac)^2 \cos n\theta d\theta ,
\end{eqnarray}
and
\begin{eqnarray}
\label{eq:cn-1}
  c_n &=& \frac{2}{\pi} \int_{0}^{\thetac}
  (\cos\theta -\cos \thetac)^3 \cos n\theta d\theta .
\end{eqnarray}

After tedious but straightforward calculations, we obtain
\begin{eqnarray}
\label{eq:b0}
\frac{b_0}{2} &=& \frac{1}{\pi}
\left(\thetac + \frac{\thetac}{2} \cos 2 \thetac
- \frac{3}{4} \sin 2 \thetac \right), \\
\label{eq:b1}
b_1 &=& \frac{2}{\pi}\left( - \thetac \cos \thetac
-\frac{3}{4}\sin\thetac + \frac{1}{12} \sin 3\thetac \right), \\
\label{eq:b2}
b_2 &=& \frac{2}{\pi}
\Big(\frac{\thetac}{4} - \frac{1}{6}\sin2\thetac
+\frac{1}{48}\sin4\thetac\Big)
,\\
\label{eq:bn}
b_n &=& \frac{1}{2n\pi} 
\Big[\frac{\sin (n-2) \thetac}{(n-1)(n-2)}
  -\frac{2 \sin n \thetac}{(n-1)(n+1)}
+\frac{\sin (n+2) \thetac}{(n+1)(n+2)} \Big]  \qquad (n \geq 3),
\end{eqnarray}
and
\begin{eqnarray}
\label{eq:c0}
\frac{c_0}{2} &=& \frac{1}{\pi}
\left(- \frac{9\thetac}{4} \cos\thetac + \frac{9}{8}\sin\thetac
- \frac{\thetac}{4} \cos 3\thetac + \frac{11}{24} \sin 3\thetac
\right), \\
\label{eq:c1}
c_1 &=& \frac{2}{\pi}  \left(\frac{9}{8} \thetac
+ \frac{3\thetac}{4} \cos 2 \thetac
- \frac{7}{8} \sin 2 \thetac -\frac{1}{32}\sin 4 \thetac
\right), \\
\label{eq:c2}
c_2 &=& \frac{2}{\pi}
\Big( - \frac{3 \thetac}{4} \cos \thetac +\frac{1}{2} \sin\thetac 
-  \frac{3}{32} \sin 3\thetac - \frac{1}{160} \sin 5\thetac
\Big), \\
\label{eq:c3}
c_3 &=& \frac{2}{\pi}
\left(\frac{1}{8} \thetac - \frac{3}{32} \sin 2 \thetac
+ \frac{3}{160} \sin 4 \thetac - \frac{1}{480}\sin 6 \thetac \right), \\
\label{eq:c4}
c_n &=& \frac{3}{4n\pi} 
\biggl[
\frac{\sin (n-3)\thetac}{(n-1)(n-2)(n-3)}
-\frac{3\sin (n-1)\thetac}{(n-1)(n+1)(n-2)}\cr
&&  \qquad\qquad + \frac{3\sin (n+1)\thetac}{(n-1)(n+1)(n+2)}
-\frac{\sin (n+3)\thetac}{(n+1)(n+2)(n+3)}
\biggr]\qquad (n \geq 4).
\end{eqnarray}
The above coefficients are combined and form the coefficients
  ${A}_n$ in equation (\ref{eq:A-Gamma<1}) of the main text:
\begin{equation}
A_n \equiv  (1 - u_1 - u_2) (\cos \ello \cos \ells) a_n + (u_1 + 2 u_2) 
(\cos \ello \cos \ells)^2 b_n - u_2 (\cos \ello \cos \ells)^3 c_n.
\end{equation}

\section{Fourier series expansion for spots
  with $\Gamma>1$ \label{sec:FT-ld-2}}

For those spots with $\Gamma>1$, the coefficients ${A}_n$ for
  spots with $|\Gamma|<1$ should be replaced by $\tilde{A}_n$, which
  are defined through
\begin{eqnarray}
\label{eq:muI}
\mu_{\rm s} I(\mu_{\rm s})
= \frac{\tilde{A}_0}{2}
+ \sum_{n=1}^{\infty} \tilde{A}_n \cos n \omegas t ,
\end{eqnarray}
where
\begin{eqnarray}
\mu_{\rm s} &=& \cos\ello \cos\ells(\cos\omegas t + \Gamma), \\
I(\mu) &=& 1-u_1(1-\mu)-u_2(1-\mu)^2 = (1-u_1-u_2) +(u_1+2u_2)\mu - u_2\mu^2.
\end{eqnarray}
 Unlike in Appendix \ref{sec:FT-ld}, the left-hand-side of equation
 (\ref{eq:muI}) is explicitly written in terms of up to the
 third-order polynomials of $\cos\omegas t$.  Thus, $\tilde{A}_n$ can
 be explicitly given in the following forms:
\begin{eqnarray}
\label{eq:A-Gamma=1}
  \frac{\tilde{A}_0}{2} &=&
   (1 - u_1 - u_2) (\cos \ello \cos \ells) \Gamma
  + (u_1 + 2 u_2) (\cos \ello \cos \ells)^2 \left(\Gamma^2+\frac{1}{2}\right)\cr
  && \qquad - u_2 (\cos \ello \cos \ells)^3
  \left(\Gamma^3+\frac{3\Gamma}{2}\right),  \\
  \tilde{A}_1 &=&
   (1 - u_1 - u_2) (\cos \ello \cos \ells) 
  + 2(u_1 + 2 u_2) (\cos \ello \cos \ells)^2 \Gamma\cr
  && \qquad - u_2 (\cos \ello \cos \ells)^3 \left(3\Gamma^2+\frac{3}{4}\right),
  \\
  \tilde{A}_2 &=&
   \frac{u_1 + 2 u_2}{2} (\cos \ello \cos \ells)^2 \Gamma
  - u_2 (\cos \ello \cos \ells)^3 \frac{3\Gamma}{2},  \\
  \tilde{A}_3 &=&
  - \frac{u_2}{4} (\cos \ello \cos \ells)^3,  \\
  \tilde{A}_n &=& 0 \qquad (n\geq 4).
\end{eqnarray}

\end{document}